\documentclass[a4paper,11pt]{article}
\usepackage{pos}

\usepackage{wrapfig}
\title{Review on Algorithms for dynamical fermions}

\author*[a,b]{Jacob Finkenrath}

\affiliation[a]{Bergische Universit\"at Wuppertal,\\
Gau{\ss}stra{\ss}e 20, 42119 Wuppertal, Germany}

\affiliation[b]{The Cyprus Institute, \\20 Konstantinou Kavafi Street, 2121 Nicosia, Cyprus}

\emailAdd{finkenrath@uni-wuppertal.de}

\abstract{
This review gives an overview on the research of algorithms for dynamical fermions used in large scale lattice QCD simulations.

First a short overview on the state-of-the-art of ensemble generation at the physical point is given.

Followed by an overview on necessary steps towards simulation of large lattices with the Hybrid Monte Carlo algorithm.
Here, the status of iterative solvers and tuning procedures for numerical integrators within the molecular dynamics are discussed.

This is followed by a review on the on-going developments for algorithms, with a focus on methods which are potentially useful to simulate gauge theories at very fine lattice spacings,
i.e.~well suited to overcome freezing of the topological charge. This includes modification of the HMC algorithm as well as a discussion of algorithms which includes the fermion weight via global correction steps.
Parts of the discussions are on the application of generative models via gauge equivariant flows 
as well as multi-level algorithms.
}

\FullConference{%
  The 39th International Symposium on Lattice Field Theory (Lattice2022),\\
  8-13 August, 2022 \\
  Bonn, Germany 
}

\begin{document}
\maketitle

\section{Introduction}

The strong interaction between quarks and gluons is described by Quantum Chromodynamics (QCD). While the theory is asymptotic free towards high energies, at low energies the coupling becomes strong and quarks are confined within color-neutral hadronic states. In the low energy regime QCD can be solved on a discretised Euclidean space-time lattice with High Performance Computing (HPC) via lattice QCD, the so-far only known \textit{ab-initio} approach. 

An observable $\mathcal{O}$ at a set of \textit{bare physics} and \textit{lattice} parameters $\{\beta, am_i,  \ldots \}$ can be evaluated via the path-integral
\begin{equation}
\langle \mathcal{O}\rangle = \frac{1}{Z} \int \mathcal{D}[U] \; \mathcal{O}(U) \cdot p(U) \qquad \textrm{with} \qquad p(U) = \left(\prod^{N_f}_i \textrm{det} D(U,m_i)\right) \cdot e^{-\beta S_g(U)} 
\label{eq:path}
\end{equation}
for $N_f$ number of quarks.
At a line of constant physics, i.e.~appropriate tuned quark masses $m_i$ and a constant physical volume,
continuum QCD can be recovered by an extrapolation in the lattice spacing $a$ by using a set of different gauge couplings $\beta$.
The integral measure is given by $\mathcal{D}[U]=\prod_{x=0,\mu=1}^{V,d} dH(U_\mu(x))$ with $dH$ the Haar measure of SU(3), $V=L^3\times T$ the lattice volume and $d=4$ the dimensions. The Boltzmann factor $p(U)$ depends on the discretised Dirac operator $D(U,m_i)$ and the pure gauge action $S_g(U)$. 

The standard method to solve eq.~\eqref{eq:path}, given by a high dimensional integral, is to use Markov Chain Monte Carlo simulations (MCMC).
The most common MCMC approaches can be roughly decomposed in two steps, 
a proposal step, where a new gauge configuration $U'$ is generated with a conditional weight $q(U)$
with the proposal probability $T(U\rightarrow U')$, and a correction step 
\begin{equation}
P_{acc} (U,U') = \textrm{min} \left[ 1, \frac{p(U') q(U)}{p(U) q(U')}\right]  ~.
\label{eq:acc}
\end{equation}
If the ratio $r(U,U')=(p(U') q(U))/(p(U) q(U'))$ is log-normal distributed, the acceptance rate \cite{Creutz:1988wv} is given by
\begin{equation}
\langle P_{acc} (U,U') \rangle = \textrm{erfc}\left\{\sqrt{\sigma^2/8}\right\} \qquad \textrm{with} \qquad \sigma^2 = \langle r(U,U')^2\rangle -  \langle r(U,U')\rangle^2~.
\label{eq:accrate}
\end{equation}
A set of $N$ configuration $U_i$ in the thermodynamical equilibrium with weight $p(U)$ is called ensemble, and if the MCMC procedure (see for more details \cite{Luscher:2010ae}) fulfils the fix-point or stability condition
\begin{equation}
 \int \mathcal{D}[U] T(U\rightarrow U') p(U) = p(U') \qquad \textrm{for all} \;\; U'~, 
\end{equation}
it follows 
\begin{equation}
\langle \mathcal{O} \rangle = \frac{1}{N} \sum_{i=1}^N \mathcal{O}(U_i) + \mathcal{O} \left( \sqrt{\frac{2 \tau_{int}}{N}} \right)
\label{eq:obs}
\end{equation}
where the autocorrelation time $\tau_{int}$ depends on the MCMC method.

Taking the limits towards continuum physics is extremely computational challenging, i.e.~roughly the computational cost increases $\propto a^{-\gamma_0} m^{-\gamma_1} L^{\gamma_2}$
with power laws $\gamma_0,\gamma_1,\gamma_2 > 1$.
A major part of the computations are the fermionic contributions, which can be represented by the inverse of the Dirac operator, e.g.~in case of the Boltzmann-factor via pseudofermions \cite{Finkenrath:2013soa}
\begin{equation}
\textrm{det} D(m_i, U) = \int \mathcal{D}[\eta] \textrm{exp}\{-\eta^\dagger D(m_i,U)^{-1} \eta \} \quad \textrm{if} \quad  x^\dagger (D+D^\dagger) x/x^\dagger x > 0 \quad \forall x \in \mathbb{C}^{12 V}~.
\label{eq:pse}
\end{equation}

\section{State of the art}

\begin{wrapfigure}[16]{L}{0.5\textwidth}
\vspace{-0.4cm}
\includegraphics[width=0.5\textwidth]{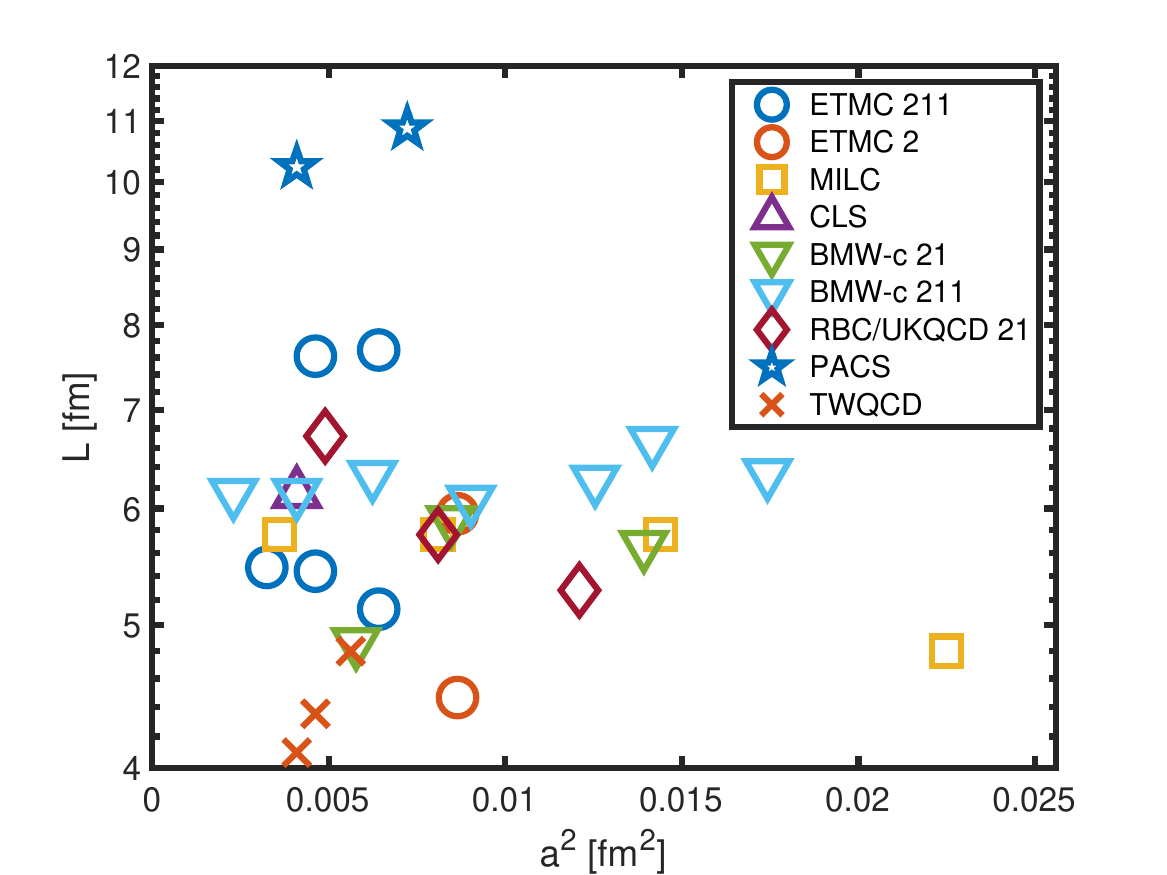}

\caption{\label{fig:ensembles}The figure shows the physical point ensembles generated by various collaborations each using different fermions discretizations (see also \cite{Bali:2022mlg}). }
\end{wrapfigure}

Due to the computational cost of computing the inverse operator $D^{-1}$, fermion determinants were neglected in the first major simulation efforts, in the so-called quenched approximation. In the early years of the millennium this led to $\sim$$10\%$ systematic effects 
in the hadron spectrum \cite{CP-PACS:2002unz}. 
This precision is not enough for advances in the low energy regime of the standard model, where search for new physics is driven by increased precision at the so-called precision frontier.
To detect signs for new physics, i.e.~by deviations with experiments, lattice QCD quantities have to be measured at sub-percentage precision, e.g.~in case of the hadronic vector contribution (HVP) to the anomalous magnetic moment of the muon.
To reach this precision, lattice QCD simulations have to include fermions in the simulation, also called using dynamical fermions, and control all major systematic effects, like finite size, finite discretization and light quark mass effects. The later effect is eliminated by directly simulating at the physical point, where the quark masses are tuned to reproduce the physical meson masses, such as pion and kaon masses. Directly simulating at the physical point is possible due to advances on the algorithmic level as well as on the hardware site.
Nowadays these physical point ensembles are generated by various lattice collaborations around the globe. 
The selected actions of the collaborations differs by the used gauge as well as the used fermion discretization, but most ensembles are generated at the isosymmetric point, e.g.~with two mass-degenerated light quarks, and
a dynamical strange and in most cases with a dynamical charm quark, denoted as $N_f=2+1(+1)$. 
Most of the generated ensembles, see Fig.~\ref{fig:ensembles} for an overview, have an lattice size of $> 5 \textrm{fm}$ and are generated at lattice spacings in the range of $[0.05 - 0.2]$ fm.
A set of ensembles in this range enables to control the major systematic effects, i.e.~finite volume and cut-off effects, in order to reach $\mathcal{O}(1\%)$ precision
in observables.


The MCMC algorithm of choice to generate these ensembles is given by the Hybrid Monte Carlo (HMC) \cite{Duane:1987de,Gottlieb:1987mq} 
algorithm. The method is used by all collaboration with different variants to improve computational efficiency, i.e.~which are based on Infra-red/ultra-violet (IR/UV) preconditioning.
The most common techniques are given by even-odd-reduction, by
Hasenbusch-mass-preconditioning \cite{Hasenbusch:2001ne} in the light quark sector and by 
rational HMC \cite{Clark:2006fx}, based on rational approximation, in the heavy quark sector. 
A subvariant used in particular for Domain Wall fermions is given by the Exact One Flavor algorithm \cite{Lippert:1999up,Chen:2014hyy},
which decompose the Dirac operator into two hermitian operators in spin space making use of the $\gamma_5$ hermicity.
Note that a detailed description on the used fermion discretizations as well as on the used preconditioning methods for the ensemble generation of the different collaborations can be found in \cite{Bali:2022mlg}.


Despite the elimination of quark mass effects, major challenges are remaining at the precision frontier. 
To make advances
the remaining systematic effects need to be further minimized to be sensible
in the search for new physics beyond the standard model.
The major contributions are given by finite size effects, e.g.~in the HVP of the anomalous magnetic moment
\cite{Borsanyi:2020mff}, and finite discretization effects present in continuum extrapolations, see e.g.~\cite{Lattice22Husung,Husung:2019ytz,Ce:2021xgd}.

\begin{wrapfigure}[15]{R}{0.5\textwidth}
\vspace{-1.8cm}
\includegraphics[width=0.5\textwidth]{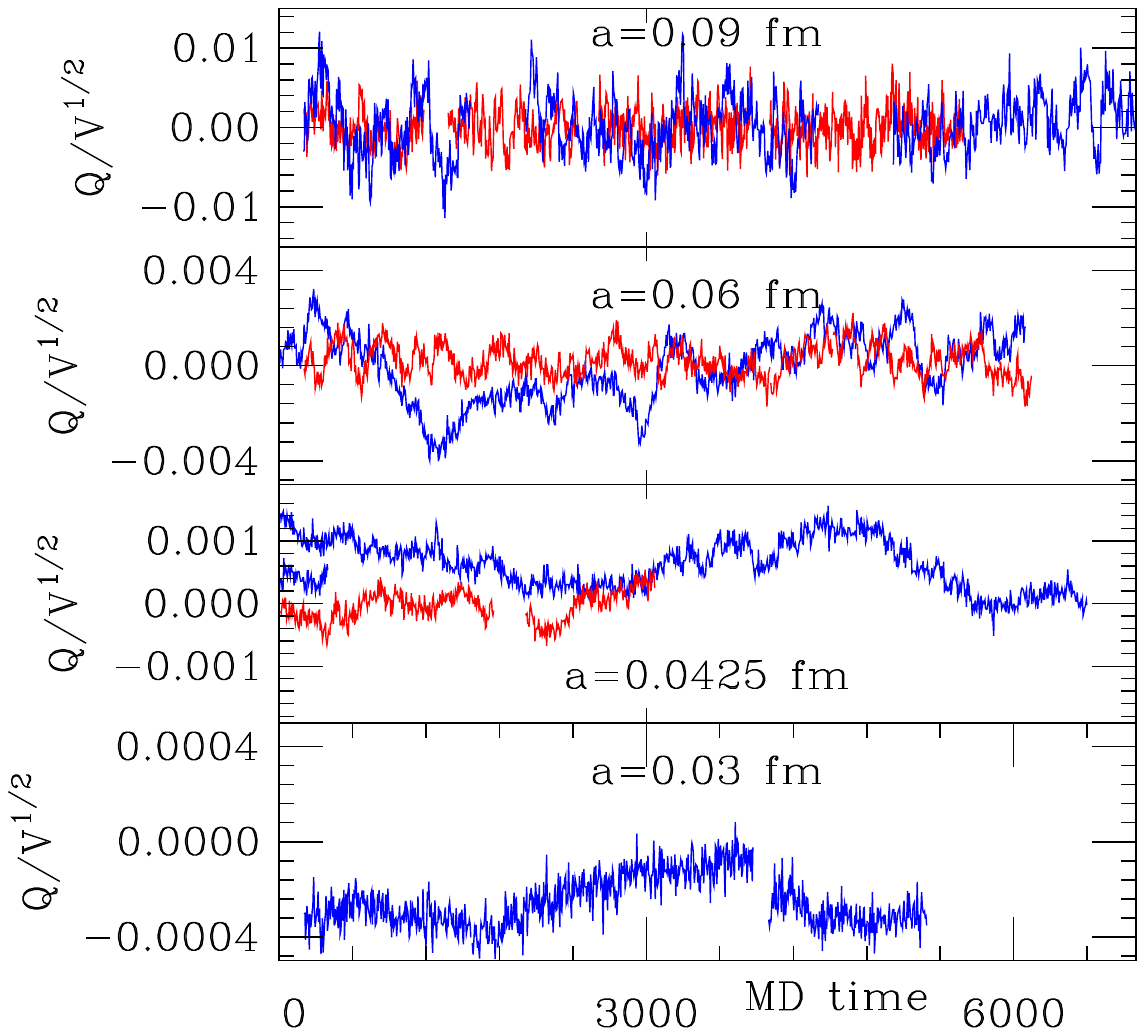}
 \vspace{-3cm}
\caption{\label{fig:autocor}Figure shows the fluctuation of the topological charge in case of the MILC $N_f=2+1+1$ HISQ staggered ensembles \protect{\cite{Bernard:2017npd}} (red history is obtained at physical pion masses). }
\end{wrapfigure}

Larger and finer lattices are needed to extend our knowledge in various directions, such as in the field of 
bottom- and charm-physics, in measurements of multiple particle scattering or in improving lattice results for the anomalous magnetic moment
g-2. The next step here is to generate new ensembles with larger lattice size, such as $> 8\;\textrm{fm}$ as well as new ensembles at finer lattice spacings $<0.05\; \textrm{fm}$.
These steps are already consider by various lattice groups, for example by the US \textit{snowmass study} in targeting ensemble generation at a fine lattice spacing of $a=0.04\;\textrm{fm}$ and a lattice extent of $L=256$ \cite{Boyle:2022uba,Boyle:2022ncb} or is already under production by the PACS collaboration at $a=0.043\;\textrm{fm}$ with $L=256$ 
\cite{Bali:2022mlg}.

However, generating these ensembles requires to meet two major challenges at the computational and algorithmic level.
For $L>128$ it requires to manage large computational costs per HMC-step at simultaneous relative large real simulation time up to 10h-24h per molecular dynamics integration.
For $a< 0.05 \;\textrm{fm}$ it requires new algorithmic advances towards finer lattice spacings because the HMC algorithm develops very large autocorrelation times due to topological charge freezing \cite{Bernard:2017npd,Schaefer:2010hu} at very fine lattice spacings, see Fig.~\ref{fig:autocor}.
This makes it practically impossible to generate an equilibrated ensemble with available computational resources below $a \ll 0.05 \;\textrm{fm}$ with periodic boundary conditions.

In this review, we will take a closer look to this two challenges. We will discuss the status of generating ensembles using the HMC algorithms in Sec.~\ref{sec:hmc}, in particular towards larger lattices. In the second part in Sec.~\ref{sec:alg} of the review, we will discuss algorithmic approaches, which can overcome limitations of the commonly used HMC algorithm, i.e.~by modifications or alternative approaches. 

\section{Scaling of Hybrid Monte Carlo}
\label{sec:hmc}

The Hybrid Monte Carlo algorithm \cite{Duane:1987de,Gottlieb:1987mq} is a MCMC algorithm, which samples configurations weighted with the Boltzmann weight $p(U)$.
This is done by iterating the following steps (see also \cite{Luscher:2010ae}) . 
\begin{itemize}
\item [1.] Generating conjugated momentas $P$ and pseudofermion via heat-bath steps. 
\item [2.] Obtaining a proposal of a new set $(U,P)$ at Monte Carlo (MC)-time $t=\tau$ by integrating Hamilton’s Equations via 
\begin{equation}
\frac{dP}{dt}= - \frac{\partial H}{\partial U} \qquad \textrm{and} \qquad \frac{dU}{dt} = \frac{\partial H}{\partial P}~.
\end{equation}
\item [3.] Followed by an acceptance step with
accept-reject probability
\begin{equation}
    P_{acc} = \textrm{min}\left[ 1, e^{-H(U_\tau,P_\tau)+H(U_0,P_0)}\right],
\end{equation}
where $H= P^{2}/2 + \phi^\dagger [D^\dagger(U) D(U)]^{-1} \phi + \beta S_g(U) $ is a basic Hamiltonian for mass-degenerated Wilson light quarks. 
\end{itemize}

The major part of the computational effort to generate an ensemble is roughly given by the size $N$, i.e.~the number of configurations, and the cost per molecular dynamics (MD) .
The cost for the MD can be hierarchical decomposed into the number of integration steps $N_{step}$ per MD  (see subsec.~\ref{subsec:int}) and the numerical cost to invert the fermion matrix $\textrm{cost}^{(inv)}$ (see subsec.~\ref{subsec:sol}).
The computational cost of the inversion can be further decomposed into linear algebra operations, which are dominated by the matrix vector product (see subsec.~\ref{subsec:bench}).
Note that the same hierarchy is introduced at the software level. Namely, the lowest level consists of linear algebraic function, the intermediate level of linear solver methods and the highest level of numerical integrators.

The challenge towards larger lattices does not only depend on the increasing computational cost, but also on how the computation can be parallelized and how it performs on novel HPC hardware.  In case of MCMC simulations the potential parallelization is limited by the sequential nature of a Markov Chain. In general this means speed up can only be achieved by parallelization at fixed lattice size, i.e.~limited by the strong scaling of the used algorithms. 
In case the upper bound of the window, where the algorithm still scales, does not scale as well as the computational cost, it results into larger simulation times even if computational resources are not limited.

Another challenge is given by the HPC architecture which comes with a relative short life cycle with new architectures entering HPC on a roughly regular two year basis while remaining at the top for roughly 6 years. This requires a constant effort in adapting the computational kernels to be performend on the updated and sometimes completely novel hardware, e.g.~on the up-coming HPC machines equipped with next-gen GPUs from different vendors such as Nvidia, AMD or Intel.






\subsection{Wilson Dirac stencil on European Supercomputers}
\label{subsec:bench}

\begin{figure}
\begin{center}
\includegraphics[width=0.9\textwidth]{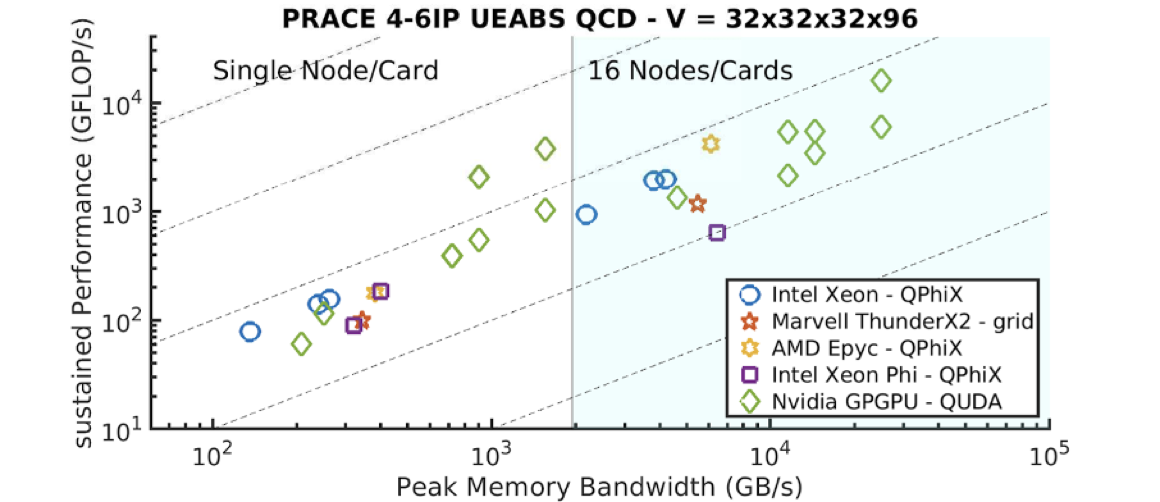}
\caption{\label{fig:PRACEEU}The figure shows the sustained performance in GFLOP/s of CG benchmarks on European HPC systems
Tier 0 obtained within PRACE 4IP-6IP on a single and 16 nodes. The two QUDA outlayers corresponds to using 4 GPUs per node.  }
\end{center}
\end{figure}

The lowest level of the lattice QCD software stack consists of linear algebra operations, such as the matrix vector product given by the Dirac stencil 
$D(U) x$. The Dirac operator $D(U)$ can be represented as a huge sparse matrix with next-neighbour interaction, which requires exchange of boundary terms if parallelised.
The arithmetic intensity of the stencil is roughly given by $1.0$
and the computation cost grows with the lattice volume $V$.
On current HPC-hardware the operator is bandwidth bound, while it becomes latency bound within the strong scaling limit.
The computational kernel, often within a conjugate gradient procedure, is used to benchmark HPC machines, e.g.~Fig.~\ref{fig:PRACEEU}
shows results from the European supercomputers by using various software packages. Namely, for Intel Xeon (Phi) architecture QPhiX \cite{QPhiX} was used, while for machines with Nvidia GPUs QUDA \cite{Clark:2009wm} and for the ARM machine the kernel from grid \cite{Boyle:2016lbp,Yamaguchi:2022feu}.
The results illustrate that the single node performance increased as well as strong scalability improved continuously over the last 10 years.

Note that performance results for novel HPC hardware are still missing, however
several lattice QCD packages already offer optimized stencil operation such as QUDA or grid.
Strong scalability for QCD kernels on Fugakus Fujitsu A64fx chips using Bridge++ can be found under \cite{Lattice22Kanamori}.
The algorithmic intensity can be reduced by introducing multiple right hand sides (rhs), while this increases the computational effort per application
 arithmetic intensity can be reduces and strong scalability as well as performance can be improved.
Several packages such as DDalphaAMG \cite{Yamamoto:2022zqp}, QUDA \cite{Clark:2017ekr} or grid \cite{Richtmann:2022fwb} are providing these kernels.


\subsection{Solvers}
\label{subsec:sol}

The integration of the MD during the HMC requires to solve frequently the Dirac equation
\begin{equation}
D(U) \cdot x = b \quad \textrm{with} \quad x,b \in \mathbb{C}^{12V} \quad \textrm{and} \quad D(U) \in \mathbb{C}^{12V \times 12V}.
\end{equation}
with known right hand side (rhs) $b$.
For large sparse matrices, like the Dirac operator $D(u)$, the common procedure is based on iterative methods which builds a Krylov space
 $\mathcal{K}_{n+1}(A,x_0) = \{x_i | i=0, \ldots, n \; \textrm{with} \;  x_i = A^{i} x_0\}$
to find the solution $x$.
The conjugate gradient (CG) solver requires a hermitian matrix, which we can derive by writing $A=D^\dagger D$,
while for more flexible methods, such as the flexible generalized minimal residual method (FGMRES), we directly use $A=D$. 
Note that in many cases, using the better conditioned even-odd reduced operator yields an effective speedup of the inversion.
For this review we will use configurations of different ensembles at different lattice sizes and lattice spacings generated by the ETM collaboration 
at physical quark masses, here referred to as cB64, cB96, cC80 and cD96 where $B:0.08 \; \textrm{fm}$, $C:0.069 \; \textrm{fm}$  and $D:0.058 \; \textrm{fm}$ and  for example cB64 corresponds to an lattice extent of $L=64$ (see also \cite{Finkenrath:2022eon,Alexandrou:2018egz}).
The ensembles will be used to discuss results obtained by the CG solver as well as for different implementation of the multigrid (MG) method.

\subsubsection{Conjugate Gradient solver}

\begin{figure}[htbp]
\begin{center}
\includegraphics[width=0.46\textwidth]{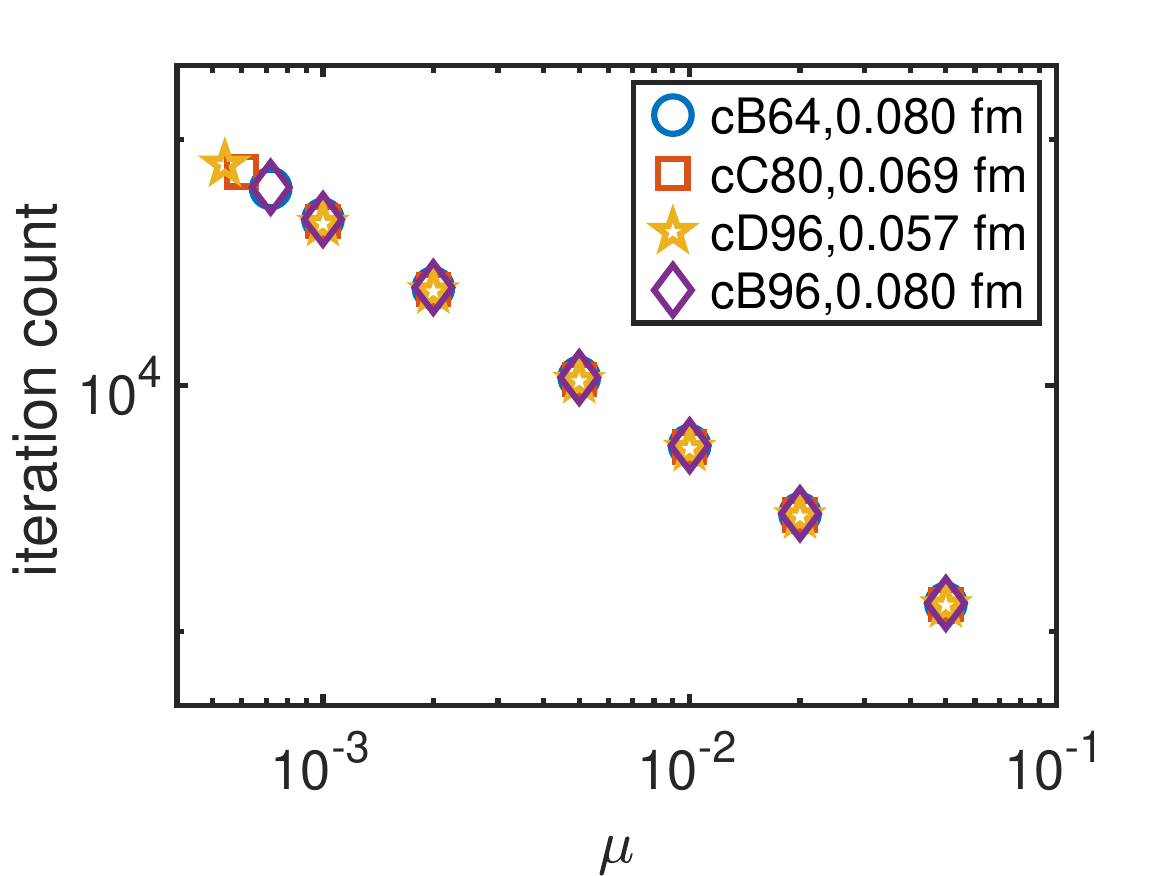}
\includegraphics[width=0.48\textwidth]{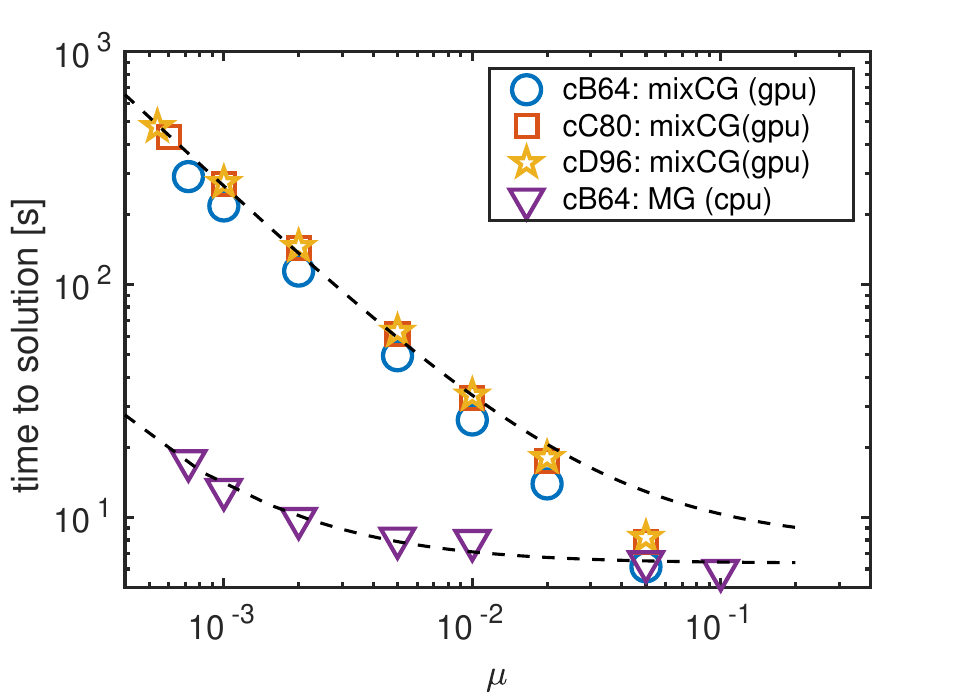}
\caption{The left panel shows the iteration per solution of the CG solver
measured on the ETMC physical point ensembles in dependence of the quark mass.
The right figure shows the time per solution obtained using the CG implemented in QUDA and MG solver implemented in DDalphaAMG.
All results were obtain using JUWELS Boosters equipped with Nvidia A100 while the CPU data is obtain on Intel Skylakes CPUs on SuperMUC.
}
\label{fig:cg}
\end{center}
\end{figure}

The most basic Krylov subspace solver is given by the conjugate gradient (CG) method,
which is commonly used for large masses,  e.g.~see Fig.~\ref{fig:cg}.
It depends mainly on the matrix-vector stencil and can be efficiently parallelized with very good scalability compared to other QCD kernels.
The algorithm can be further sped up by using mix-precision methods, i.e.~performing most iteration using low-precision arithmetics.
This leads to a speed up by $40\%$ using single and $50\%$ half precision on Juwels Booster A100 GPUs, as shown on the left panel in Fig.~\ref{fig:cg} . As pointed out by \cite{Lattice22Clark}
this can be further optimised using a suitable representation of low-precision numbers.

The number of needed iterations of the CG solver is proportional to the condition number of the operator, i.e.~in case of lattice QCD this reduces to
the dependence on the smallest eigenvalue. It is independent from the density.
Towards the physical point the iteration count drastically increases towards $~70k$ iteration.
For the computational costs we found
\begin{equation}
cost_{CG} \approx V \cdot \left( \frac{b}{\mu} + a \right) \approx V \frac{b}{\mu}
\end{equation}
with $b/a \sim 0.04$ as depicted on the right panel in Fig.~\ref{fig:cg}.

\subsubsection{Multigrid approaches}


The high iteration count of the CG solver at the physical light quark mass can be overcome by using preconditioned Krylov subspace solvers.
A very effective method in reducing the iteration count to $\mathcal{O}(10)$ is given by algebraic multigrid (MG) procedures.
This is done by using a flexible solver such as FGMRES and treating the highly fluctuating UV modes via a smoother, e.g.~based on Gauss-Seidel iterations or on a Schwarz-alternating procedure, and the low fluctuating IR modes via a coarse grid correction based on an algebraic multigrid approach.

Multigrid procedures outperform the basic CG methods by up to two orders of magnitude for several versions of lattice fermion actions. Examples are  Wilson fermions
\cite{Luscher:2007es,Luscher:2007se,Babich:2010qb,Frommer:2013fsa} and twisted mass fermions \cite{Alexandrou:2016izb}. While for other fermion types
like staggered fermions \cite{Brower:2018ymy,Lattice22Ayyar} or Domain Wall fermions \cite{Brower:2020xmc,Boyle:2021wcf} effective MG methods became 
only recently known and their improvements are less significant (up to one order of magnitude). Note that a MG procedure comes with an additional overhead, given by the setup time for building the coarse grid projection and restriction operator. 

\begin{figure}
\begin{center}
\includegraphics[width=0.47\textwidth]{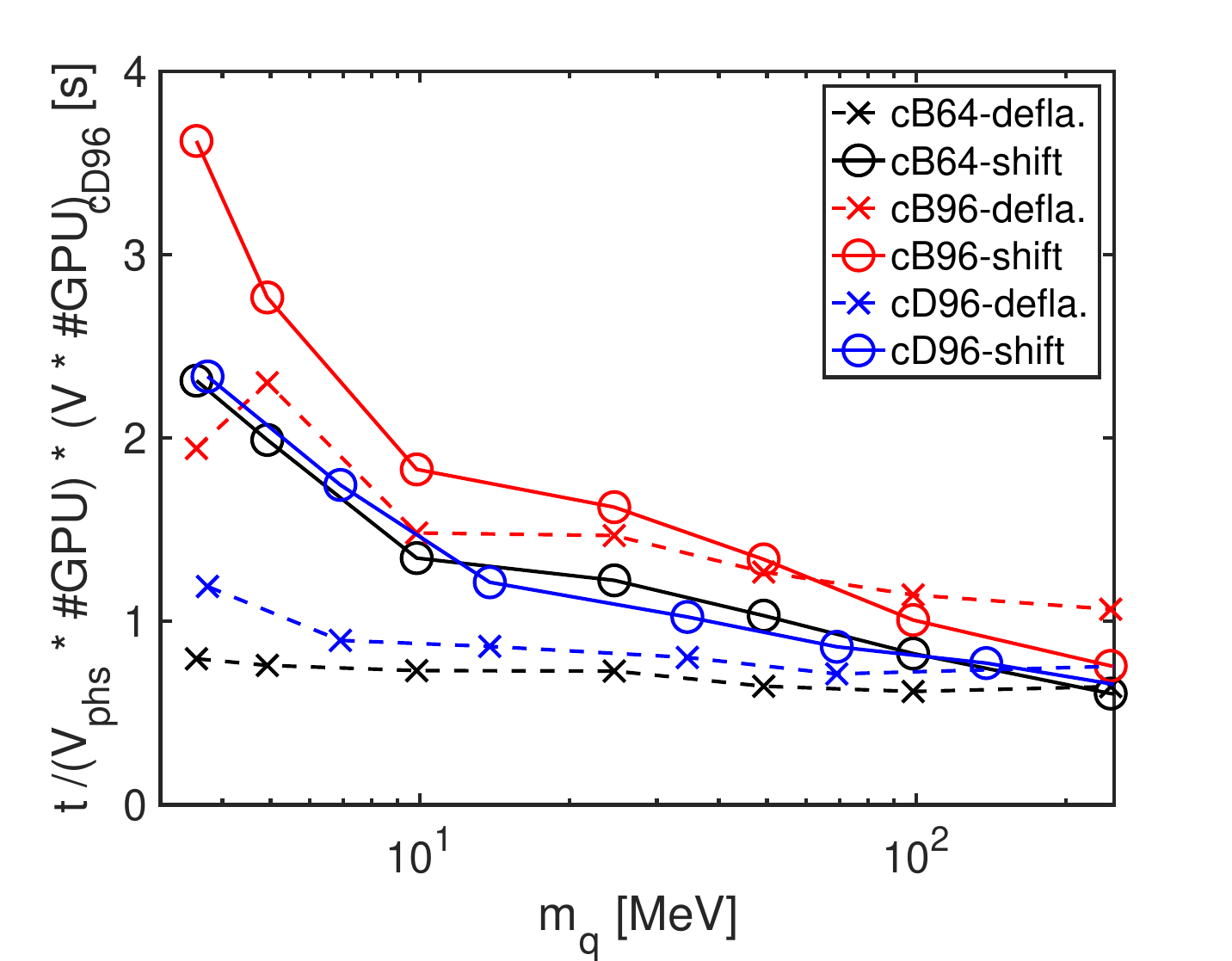}
\includegraphics[width=0.49\textwidth]{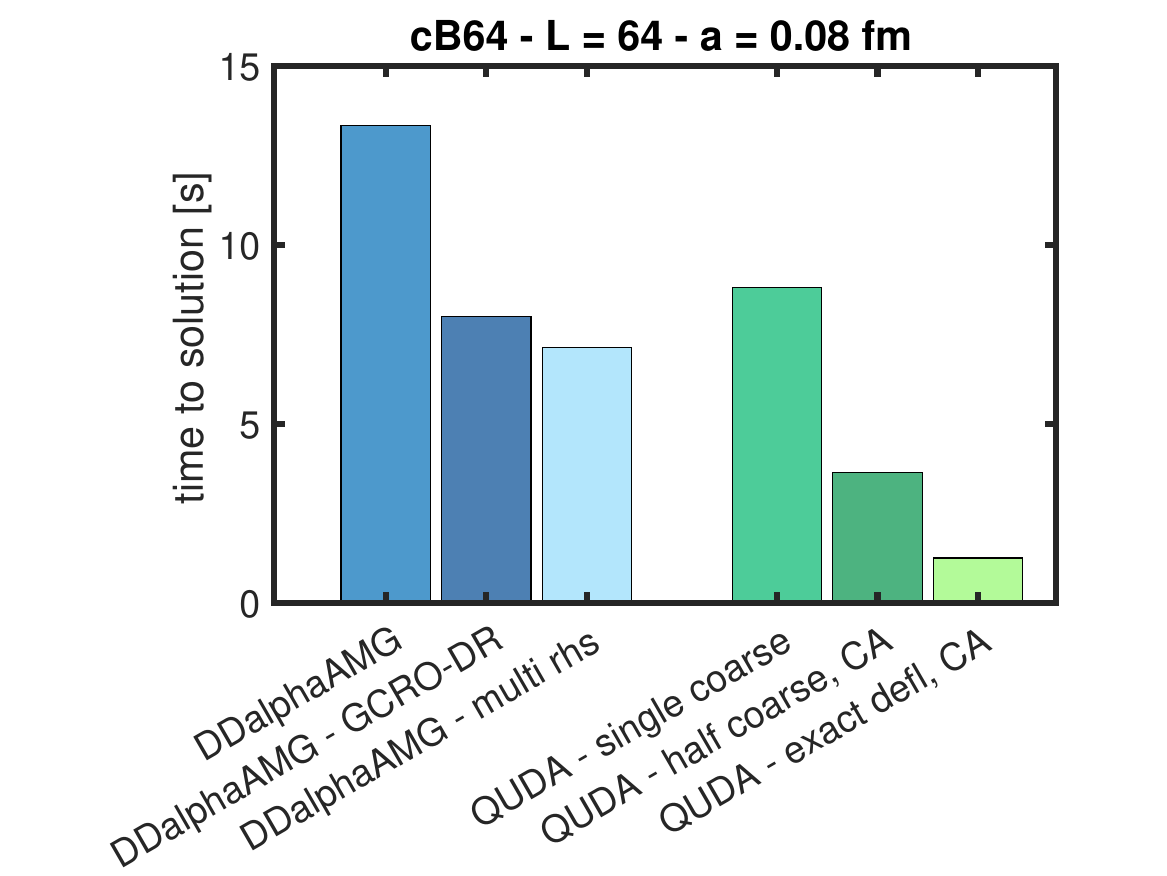}
\caption{The left figure shows the mass-dependence of a 3 level MG solver implemented in QUDA using a coarse grid shift in $\mu$ (solid lines) and exact deflation (dotted line) relative to the solution time on cD96 at $m_q\sim 250$ MeV. The right figure shows recent performance improvements of the DDalphaAMG solver (in blue) and of the QUDA MG solver (in green) }
\label{MG:improvements}
\end{center}
\end{figure}

As shown on the right panel of Fig.~\ref{fig:cg} MG methods are suppressing the quark mass dependence. We found for the DDalphaAMG solver $ \propto V (0.001/\mu +1)$  \cite{Alexandrou:2016izb}. 
A similar behaviour is found in case of the MG solver implemented by QUDA, see left panel of Fig.~\ref{MG:improvements}. Here, the quark mass dependence can be even further suppressed by using exact deflation on the coarsest grid.
In contrast to the CG solver, we found that the numerical costs increases more than linear with the volume at constant lattice spacing, i.e.~comparing cB64 to cB96, while keeping the physical volume constant the cost scales as expected, i.e.~comparing cB64 to cD96.  A possible explanation is that the cost scales additionally with the density of the low eigenmodes.


If MG approaches are used within the MD of the HMC, the constant update of the setup limits the effectiveness of the solvers.
The overhead can be limited by updating the previous setup with less setup iterations. This leads, however, to additional reversibility violations,
which can be minimise with higher residual precision. 
Moreover the window, where the multigrid approach scales strongly, is limited, towards large parallelization by the coarse grid size
and towards low parallelization by the memory. 

While the usage of a MG method can improve the total cost per trajectory, as shown in Fig.~\ref{fig:impr}, the limitation given for the maximal parallelization
increases the simulation time, i.e.~we found that the upper bound of the strong scaling window increases roughly with $L^3$ while the cost for the HMC increases with $L^{4.5}$. This leads to time per trajectories of around 6-8h on $L=96$ lattices on 384 Skylake nodes of SuperMUC-NG.

However this can be improved by developing more efficient coarse grid algorithm or utilizing GPU machines.
Examples for algorithmic improvements are given in case of DDalphaAMG by a multiple rhs version, see \cite{Yamamoto:2022zqp},
and by additional coarse grid improvements \cite{Espinoza-Valverde:2022pci}, based on pipelined polynominal preconditioning of a deflated restart GCR method.
The results are shown on the right panel of Fig.~\ref{MG:improvements}. 
An even more impressive improvement is seen in case of the performance improvements of the QUDA MG solver in the past years, see right panel of Fig.~\ref{MG:improvements}. Using communicating avoiding CG in combination with reduced precision, the method improved the time to solution from 8 sec on 64 Nvidia P100 nodes to 3.7 secs on 16 Nvidia A100. Deflating additional 800 exact coarse grid eigenvalues further reduces inversion times to 1.26 secs.
This also reduces significantly the quark mass dependence, as shown on the left panel of Fig.~\ref{MG:improvements}.
Moreover further improvements are expected by using multiple rhs, e.g.~shown in case of grid \cite{Boyle:2016lbp,Yamaguchi:2022feu} on Juwels Booster, see \cite{Richtmann:2022fwb}, which found significant speed up.

While improvements of the coarse grid directly improve observable calculations, MG improvements, which comes with an additional overhead, might be not directly useful for the MD integration, e.g.~exact deflation of QUDA MG procedure would require a relative large setup time such that the standard variant is significant faster.

\subsection{Integrators}
\label{subsec:int}
The HMC algorithm propose a new gauge configuration via MD integration, 
where the total cost of the HMC algorithm
is directly proportional to the number of required integration steps $N_{step}$.
The integration can be done via a symplectic, reversible integrator as required by the fix-point condition.
The most common integrator approach is given by the second minimal norm scheme \cite{Sexton:1992nu}, 
while a large collection of higher order schemes can be found in the work by Omelyan, Mryglod and Folk 
\cite{Omelyan:2001}, which also includes force gradient approaches.

To correct for the error of the numerical integration, an accept-reject step is done after each trajectory.
The acceptance rate of the accept-reject step is given by eq.~\eqref{eq:accrate}, here $\langle P_{acc} \rangle = \textrm{erfc}(\sqrt{\sigma^2/8})$
and the related variance $\sigma^2(N_{steps},\mu, V,\ldots) = var (\delta H)$ can be written as a function of integrator as well as lattice parameters \cite{Creutz:1988wv,Takaishi:2005tz,Kennedy:2012gk}. For a $n$-th order integrator, we find
\begin{equation}
 \textrm{var} (\delta H) = \frac{1}{N_{steps}^{2 n}} \sigma^2(1,m, V ,\ldots) \propto h^{2 n}  \, V \, m^{-2 \alpha_0}
 \label{eq:vardH}
\end{equation}
with $\alpha_0 > 1$ and $h=1/N_{step}$ .
Now, a suitable tuning condition is given by fixing the acceptance rate $\langle P_{acc} \rangle$ and minimizing the cost function which depends on the number of steps and the cost per force computation.
Using IR/UV preconditioning techniques, such as Hasenbusch mass preconditioning or rational approximation, can further reduce the cost but increases the search space. 

A suitable approach to calculate $\textrm{var} (\delta H)$ is given the first error terms,
which can be calculated via Poisson Brackets and are the first non-zero higher order terms
of the Shadow-Hamiltonian \cite{Kennedy:2012gk}.
In case of the second minimal norm scheme, the Shadow-Hamiltonian is given by
\begin{equation} 
\tilde{H}=T+S+h^2 \left( \frac{6\lambda^2-6 \lambda + 1}{12} \{S,\{S,T\}\} + \frac{1-6\lambda}{24} \{T,\{S,T\}\} \right)+ \mathcal{O}(h^4)~.
\end{equation}
with the freely select-able parameter $\lambda$, which roughly minimizes the $\mathcal{O}(h^2)$-term by setting to $\lambda \sim 0.19$.
If we set $\lambda=1/6$ the leading order term is given by the force $\{S,\{S,T\}\}= \textrm{tr} (F^2)/a$.
This relation can be now used to tune the step size $h$.

\begin{wrapfigure}[15]{R}{0.4\textwidth}
\includegraphics[width=0.4\textwidth]{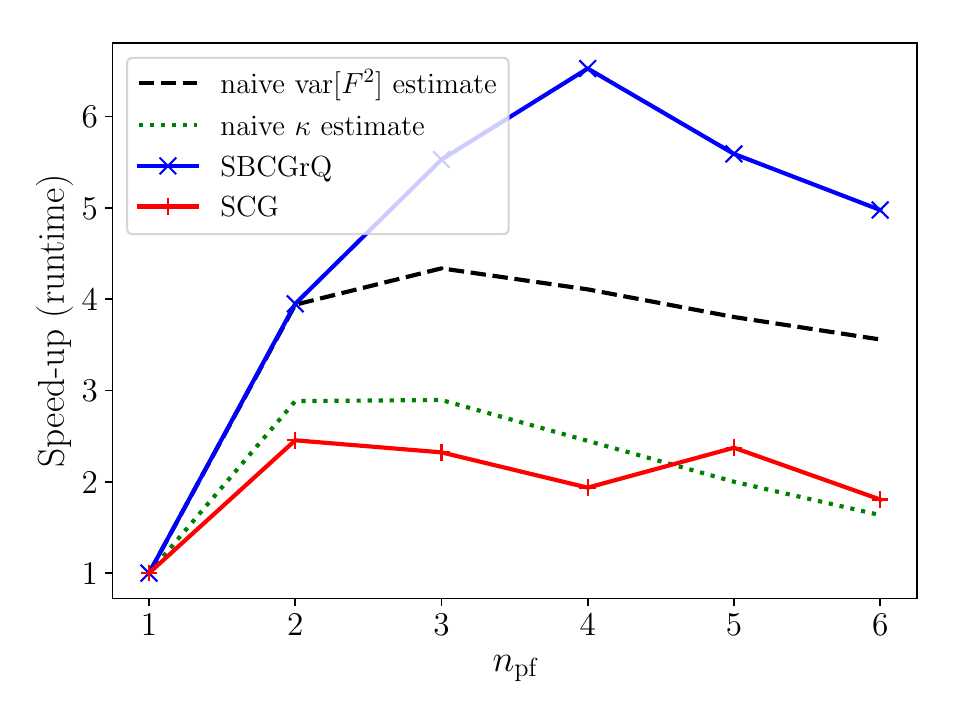}
\caption{\label{fig:BCG}The figure shows the relative speed up using higher roots $n_{pf}$ in combination with a Block Krylov solver (taken from \protect{\cite{deForcrand:2018orx}}).}
\end{wrapfigure}

Moreover, the measurement of the force $f=\partial H/\partial U$, 
can be used to evaluate methods. An example is given by the 
RHMC \cite{Clark:2006fx} with Block Krylov solvers and multiple pseudofermions \cite{deForcrand:2018orx}.
Here, a variant of RHMC is introduced by splitting up
$\textrm{det}[D^\dagger D] = \textrm{det}[(D^\dagger D)^{1/n_{pf}}]^{n_{pf}}$
and for the variance of the force follows \cite{deForcrand:2018orx}
$\textrm{var}(F^2(n_{pf})) = c_s n_{pf}^{-1} + c_3 n_{pf}^{-2} + \mathcal{O}(n_{pf}^{-3})$.
This reduces the required steps at given acceptance
and is ideal to combine with Block Krylov solvers. 
An example are given by the shifted Block CGrQ method, \cite{LKeegan:2018orx},
which leads to faster convergence by increasing the search space and is, in general, ideally to combine with multiple rhs kernels, e.g.~\cite{Yamamoto:2022zqp}.
As depicted in Fig.~\ref{fig:BCG} the RHMC combined with a Block solver leads to speed ups of $\sim 6$ achieved
in case of $N_f=4$ on a $L=8$ lattice.

IR/UV splitting of the action are naturally combined with nested integration schemes, where the high fluctuating UV terms can be integrated with a smaller step size and expensive IR terms with a larger step size.
Let us write the Hamiltonian as $H= S_1 + S_0 + P $, with $S_1$ containing the IR modes and $S_0$ the UV modes. 
Then, a nested integration scheme based on the second minimal norm scheme with $\lambda=1/6$ can be written as
\cite{Urbach:2005ji,Shcherbakov:2015hhg}
\begin{equation}
\Delta(h) = e^{\frac{h}{6}\hat{S_1}} \Delta(h/2)  e^{\frac{2h}{3}\hat{S_1}-\frac{h^3}{72}\hat{C_1}} \Delta(h/2) e^{\frac{h}{6}\hat{S_1}}
\label{eq:forcegrad}
\end{equation}
with
$\Delta(h/2) = e^{\frac{h}{12}\hat{S_0}} e^{\frac{h}{4}\hat{P}} e^{\frac{h}{3}\hat{S_0}}e^{\frac{h}{4}\hat{P}}e^{\frac{h}{12}\hat{S_0}}$
and 
$C_1 = 2 \sum^{V,3}_{x=1,\nu=0} \frac{\partial S_1 }{\partial U_{\nu}(x)} \frac{\partial^2 S_1}{\partial U_{\nu}(x)\partial U_{\mu}(x)}$
the force gradient term,
which comes with a second derivative.
If the integration errors from $S_0$ are sufficiently suppressed, this leads to an fourth order integrator.
As noted by Lin and Mawhinney \cite{Yin:2011np}, the force gradient term can be approximated numerically,
which avoids the implementation of second derivative terms. 
The approximated variant can be implemented based on the force terms and only requires more memory for an additional gaugefield.
Moreover the total cost of inversions of the fermion matrix is even reduces by 4/3 compared to the exact force gradient term, see \cite{Shcherbakov:2015hhg} .

A very common IR/UV-spliting method is given by Hasenbusch mass preconditioning \cite{Hasenbusch:2001ne}, which introduces additional mass terms $\mu_i$.
In principle, these additional parameters can be now tuned, if the first Poisson-Brackets are known. In case of a fourth order integrators like eq.~\ref{eq:forcegrad}
the brackets requires the calculation of derivative of up to the fourth order \cite{Kennedy:2012gk}.  
In practice the shifts $\mu_i$ are tuned empirically. We found roughly a dependence of 
$\propto (\frac{\Delta^2 m_i}{\mu_i^2})^k$ with $\Delta^2 m_{i} = \mu_{i+1}^2 - \mu_{i}^2 $ and $k \in [3, 4]$ for corresponding contributions to the variance of $\delta H$ eq.~\ref{eq:forcegrad}.
This leads to a relative shallow nested integration setup by using a fourth order scheme with deepth of three or even only two levels with Hasenbusch mass shifts roughly given by $\mu_i \in \{ \mu+ \mu [ 0, 1, 10, 100,\ldots ]\} $.
Note that this choice also suppresses effectively the quark mass dependence, i.e.~$\mathcal{O}(\Delta^2 m_0) \approx \mathcal{O}(\mu^2)$.

\subsubsection{HMC on GPUs}

The relative long (real) time per MD of up to 8 hours can be sped up by porting the major computational kernels to novel HPC architecture or to accelerator cards, such as GPUs.
For GPUs, highly optimized software packages like QUDA \cite{Clark:2009wm} or grid \cite{Boyle:2016lbp,Yamaguchi:2022feu} exists which can be used to offload major computational tasks, like the inversions of the Dirac operator. QUDA is already used in several frameworks, e.g.~Chroma as well as the packages MILC and CPT utilize solvers implemented in QUDA to speed up HMC-simulations, see right panel of Fig.~\ref{fig:impr}. A similar effort is on-going for twisted mass fermion implemented within tmLQCD \cite{Kostrzewa:2022hsv}.
Moreover also fully-fledged software packages implements HMC on GPUs, such as grid.

The highly optimized kernels of these package are accessible through python APIs, which enable to interact and use computational kernels in an high level environment, simplifying programming and pushing algorithmic developments. 
Implementations of the HMC method are available in the python package GPT \cite{Lattice22Lehner}, which uses grid \cite{Boyle:2016lbp,Yamaguchi:2022feu} and in the python package lyncs-API \cite{Lattice22Bacchio,Lattice22Yamamoto}, which utilises QUDA functions \cite{Clark:2009wm} to perform on GPUs.


\subsection{Conclusion}

\begin{figure}
\includegraphics[width=0.47\textwidth]{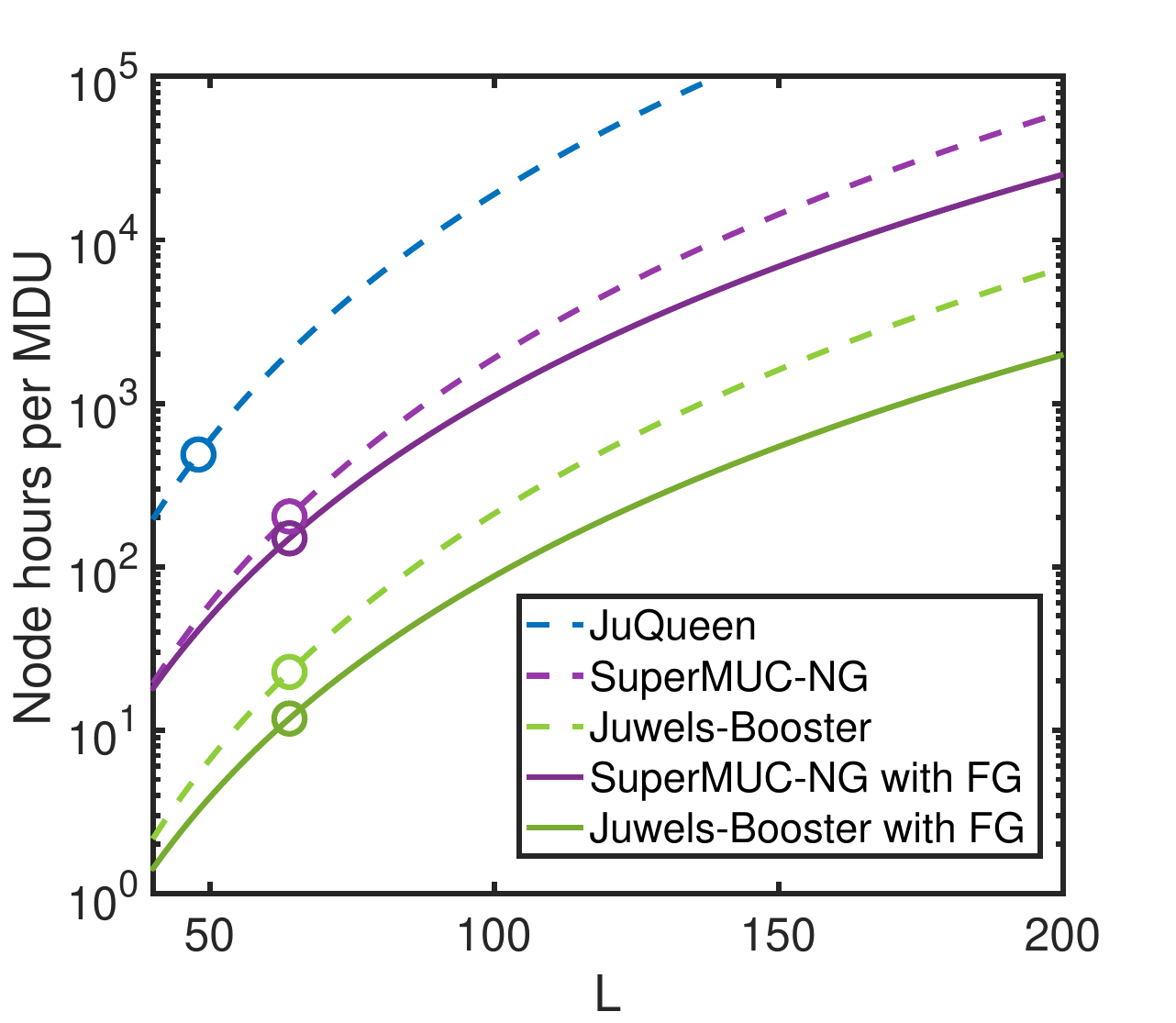}
\includegraphics[width=0.53\textwidth]{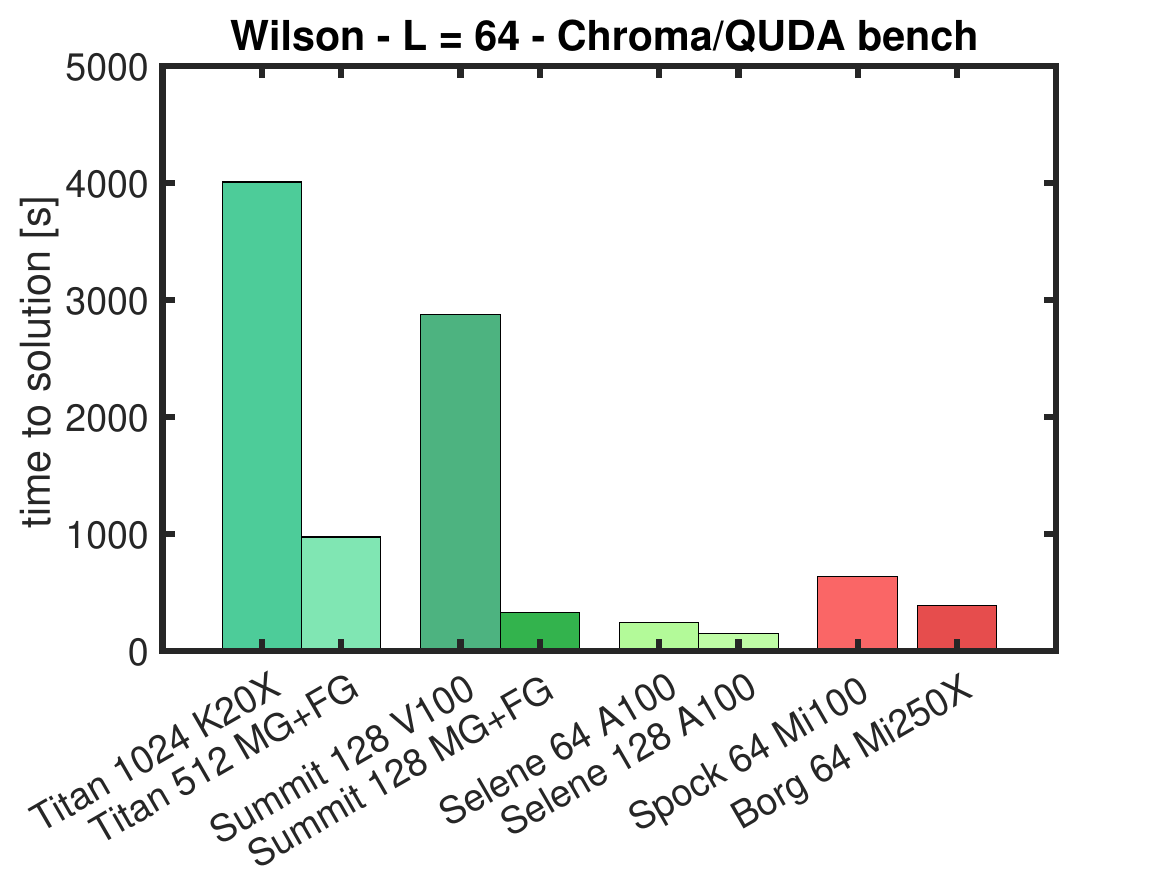}
\caption{\label{fig:impr}The figure shows the improvements on computational costs of HMC simulation of the last years. On the left side, the improvements of node hours per MDU are shown in case of tmLQCD illustrating the effect using MG solvers, higher order force gradient (FG) integrators and GPUs. On the right side the improvement time to solution of a MDU in case of using Chroma with QUDA is shown.}
\end{figure}

The computational cost for the HMC algorithms could be reduced by advances based on algorithms and hardware,
such as the usage of MG methods or the utilization of HPC machines equipped with GPUs.
In case of twisted mass fermions this leads to a reduction of node hours per trajectory in simulations of physical point ensembles.
As illustrated on the left panel of Fig.~\ref{fig:impr}, using a MG solver in combination with Intel Xeon Server CPUs gave a speed-up
by a factor 10 compare to the cost required on JuQueen. By utilizing the QUDA solvers another reduction by a factor 10 in node hours
were obtained.
A comparable improvement is found in case of Wilson fermions using a Chroma driven HMC, which utilizes QUDA. As shown on the right 
panel of Fig.~\ref{fig:impr}, a speed-up of a factor 40 was obtained using 128 Nvidia Ampere 100 of Selene compare to 1024 
Titan nodes each equipped with a Nvidia K20X GPU.

The total costs for ensemble generation by the HMC algorithm using $N_H$ Hasenbusch terms can be written as
\begin{equation}
 \textrm{cost}_{HMC} = N(\tau_{int})  \sum_{i=1}^{N_H} N_{steps,i} \times \textrm{cost}_{i}^{(inv)}  \propto  a^{-\gamma} 
  \frac{V^{1/2n}}{a^{2/n}} \frac{V}{a^4} =  a^{-(4+2/n+\gamma)} V^{4+1/2n}
\end{equation}
if we assume $\tau_{int} \propto a^{-\gamma}$ where $\gamma \approx 5$ for periodic boundaries \cite{Schaefer:2010hu}.
The dependence on the quark mass is reduced by using MG solvers and Hasenbusch mass preconditioning. 
The volume scaling can be reduced by using integrators of higher order $n$. 
For a lattice with extent $L=192$ follows that a trajectory requires  $\mathcal{O}(1000)$ node hours on a HPC machine equipped with Nvidia A100.
If strong scalability of the solvers towards this lattice sizes does not deteriorate, large lattice sizes  are in reach with exascale computing (see Fig.~\ref{fig:impr}).

\section{Towards fine lattices}
\label{sec:alg}

Towards fine lattice spacings $a<0.05 \; \textrm{fm}$ the autocorrelation time of observables calculated on configuration sampled with the HMC algorithm increases drastically with $\tau_{int} \propto a^{-5}$ or even exponentially \cite{Schaefer:2010hu}, which leads to freezing of the topological charge (see Fig.~\ref{fig:autocor}).
The phenomenon is related to the gauge group and is also present in case of pure gauge simulations.
Overcoming this critical slowing down of the MCMC algorithms towards the continuum is under active research and several different approaches are investigated.
These different methods  can be roughly categorised into two subcategories,
methods not based on the HMC proposal by changing the proposal procedure $T(U\rightarrow U')$ and including corrections via an accept-reject step,
and methods which adapt the HMC either by modifying the procedure of the MD integration or by changing the action or boundary conditions.

The major obstacle for developing efficient methods is given by the acceptance rate eq.~\eqref{eq:accrate}. If the new configuration is not proposed with care, the variance of the ratio $\textrm{var}(r(U,U'))$ is  proportional to the volume and the acceptance rate is decreases $\propto \textrm{exp}(- V)$. Only gauge proposals which can suppress this volume fluctuations will work as MCMC methods in the large volume limit. 
In general, methods based on MD integration have a very solid advantage. Namely the fluctuations only increase proportionally to the discretization error of the numerical integration $\propto V^{1/2n}$.


\subsection{Modification of the HMC}

To keep the volume scaling of the MD integration for the gauge proposal, different modification of the HMC algorithms are considered.

One possible way is to change the boundary conditions.
A common approach to overcome topological freezing is given by 
open boundary conditions in time \cite{Luscher:2011kk}. The boundary conditions allow transitions between topological sectors and reduces the corresponding autocorrelation time to $\tau_{int}(Q) \propto a^{-2}$.  This comes with the breaking of translational invariance and larger extents in the lattice time direction. 
Open boundaries are the standard in simulations of CLS and are used in simulations of very fine lattice spacings, see e.g.~\cite{Cali:2021xwh}.

Another approach is given by generating masterfields
\cite{Luscher:2017cjh,Albandea:2021lvl}. Here, the idea is to increase the physical volume such, that local topological charge fluctuations are sufficient even if the global charge is fixed.
An algorithm, which is investigated to enable this masterfield simulation, is based on stochastic molecular dynamics \cite{Lattice22Fritzsch}.
Another way is given by fixing the simulation to each topological sector and averaging over the different sectors during the calculation of the observables
\cite{Brower:2003yx,Czaban:2013haa}.
Another class of algorithms, which are using coarser and smaller lattices to propose a finer lattice is given by multiscale equilibration
\cite{Detmold:2016rnh,Detmold:2018zgk,Tu:2018dws}. However, here the coarse grid to fine grid transition requires re-thermalization of the fine grid.

Another possible way is to modify the MD integration.
For example by using skewed detailed balance \cite{Cossu:2017eys,Lattice22PintoBarros}.
One possibility is to extend the accept-reject procedure, i.e.~after a rejection the MD integration is continued, not re-started, and the additional
accept-reject is modified including the probability distribution from the initial, the first and second step \cite{Cossu:2017eys}. This can be iterated, however acceptance of the second or even higher step might be small and the procedure works likely only if the integration errors oscillate and does not grow.
Another way is to introduce transitions in an extended sampling space. This enables to include additional features, such as the change of topology, within the transition probability. This is studied by \cite{Lattice22PintoBarros} in case of the 1D O(2) model, where a skewness function is used to modify the local accept-reject steps. 
The extensions can be done, such that the fix-point condition is still fulfilled but the procedure still needs further investigations to proof effectiveness.
 
Another way is given by using trivializing maps. The idea is to map the theory 
through a variable change to a trivial theory.
Here, slow modes, e.g.~which couple to the topologocal charge, are mixed with high frequent modes and gauge updates decouples easily in few integration steps. 
However, a major task is to find an appropriated map to project between the target and trivial region.
In the initial proposal, the Wilson flow was proposed \cite{Luscher:2009eq}.  Alternative approaches to design such maps are motivated by the Schwinger-Dyson equation \cite{Lattice22Matsumoto}
or by combination with normalizing flows \cite{Albandea:2022fky,Foreman:2021ljl}.

Another possibility is to change how different modes are integrated during the MD integration.
This can be done by redefining the conjugated momenta term by using Riemannian-mannifolds or Fourier acceleration \cite{Nguyen:2021zgx}.
This effectively changes the integration steps for the low and high modes, by accelerating the integration in the direction of the slow modes and deaccelerating
the high modes.

To tame or trigger dynamics during the integration one can add marginal terms to the Hamiltonian.
An example is to add Pauli-Villars fields \cite{Hasenfratz:2021zsl}. These fields can reduce large cut-off effects in many fermion simulation such as $N_f=8,12$
and make simulations also at coarser lattice spacings possible.	

In a similar way additional terms can be added to the Hamiltonian, e.g.~which can de-correlate modes. For example, 
a metapotential, which couples to the topological charge, can be added which effectively decreases topological barriers between sectors \cite{Eichhorn:2021ccz,Eichhorn:2022wxn}. The additional terms are coming with an additional weight, which can be corrected after sampling the configurations, e.g.~by reweighting. This can indeed minimize autocorrelation times, as it was found by \cite{Eichhorn:2022wxn}, but the generation and evaluation of the metapotential comes with additional numerical costs.

Another way is to use modified integrators, such as machine learned leap frog integrators \cite{Foreman:2021rhs} to de-correlate the update.
This has to be done with some care otherwise the error of the integration might recover the linear volume behaviour.

\vspace{0.2cm}

\subsection{Global Corrections}

\begin{wrapfigure}[12]{R}{0.5\textwidth}
\vspace{-0.6cm}
\includegraphics[width=0.5\textwidth]{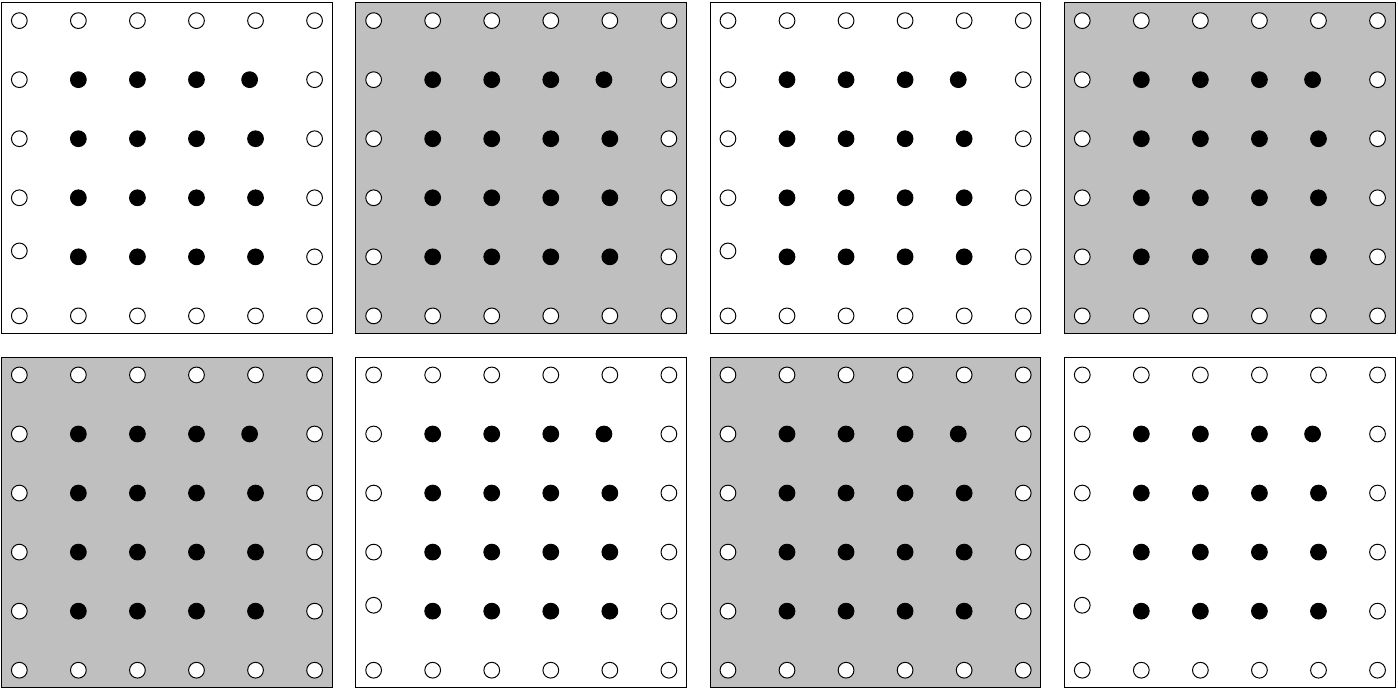}
\caption{\label{fig:decom}Domain decomposition of a two dimensional lattice into black and white blocks. 
All filled points contains links only located in the corresponding domains (taken from \cite{Luscher:2004pav}).}
\end{wrapfigure}

An obvious idea to overcome long autocorrelation times is to introduce a gauge proposal without autocorrelations.
If the distribution of the proposal is known, a combination with an accept-reject step leads to a valid MCMC algorithm, as long as the fix-point condition is satisfied.
However, as discussed, this naively leads to an ineffective algorithm at large volumes $V$, i.e.~it scales with $\langle P_{acc} \rangle \propto e^{-V}$.

The challenge is, if a suitable proposal with short autocorrelation time is found, to control the variance
of the logarithms of the ratios $r(U,U')$
\begin{equation}
\Delta S = \textrm{ln} p(U')  - \textrm{ln} p(U) - \textrm{ln}   q(U')  + \textrm{ln}  q(U) ~.
\label{eq:deltaS}
\end{equation} 
This can be done by restricting the phase space of the distributions.
Namely, the Boltzmann-factor can be factorized into a product of distributions, each depending on a
different parameter space. 

For ultra local actions, such as the pure gauge plaquette action $\beta S_g = \beta/(2N)  \sum_{x; \mu<\nu}  \textrm{Re} (1 - \textrm{tr} \, P_{\mu,\nu}(x))$, this can be done via domain 
decomposition of the lattice, e.g.~see Fig.~\ref{fig:decom}.  Then the action splits up into parts, which are independent of each other and defined within a domain 
and parts which are connecting domains, i.e.~the global distribution splits up into a product of distributions which scales with the domain sizes.
An idea is to find an transition probability, which can propose new gauge fields within a block and allows to flip the topological charge, see subsec.~\ref{subsec:gupTP}.

In case of the fermion determinant, decomposition into only local parts is not possible, however one can use the Schur decomposition
of the determinant \cite{Luscher:2004pav,Finkenrath:2012az}
\begin{equation}
\textrm{det} D = \textrm{det} S  \prod_i \textrm{det} D^{(blk)}_i \quad \textrm{with} \quad S = D_{w,w}^{(blk)} - D_{w,b} (D_{b,b}^{(blk)})^{-1} D_{b,w}
\end{equation}
which factorizes the determinant in local part, given by the block determinants, and a global part, given by the Schur compliment $S$.
By further decomposing the blocks into smaller blocks, the method becomes recursive.
The decomposition can be done via asymmetric domains using  red-black coloring, which leads to a decomposition in time used in a fermionic multi-level
approach \cite{Ce:2016idq,Ce:2016ajy}, see subsec.~\ref{subsec:mlvl} . This can be extented to a decomposition in four dimension, as discussed in \cite{Giusti:2022xdh,Saccardi:2022lzd}.
Alternative decompostions of the determinants are proposed, such as a
complete factorization in time \cite{Lattice22Wenger}.

Another possibility to control the variance of eq.~\eqref{eq:deltaS} is to use correlations between the distributions or the corresponding action.
This can done by using parameter, e.g.~via a linear parameter, which introduces shifts in the gauge coupling $\delta \beta$
\cite{Irving:1996bm,Finkenrath:2012az,Hasenbusch:2018iuw}.
This can be extented arbitrarily to a full parametrisation of the corresponding distribution.
For example one can introduce neutral networks, which can be trained via machine learning
 \cite{Albergo:2019eim}, see subsec.~\ref{subsec:generative}.

\newpage

\subsubsection{Multilevel algorithms}
\label{subsec:mlvl}

\begin{wrapfigure}[10]{L}{0.5\textwidth}
\includegraphics[width=0.5\textwidth]{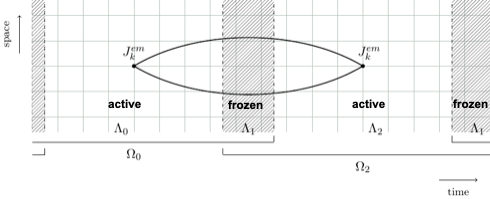}
\caption{\label{fig:MLdecomposition}The figure shows the one dimensional decompostion in time used for the multilevel algorithm used in \cite{DallaBrida:2020cik}.}
\end{wrapfigure}

Domain decomposition techniques are relevant for other aspects. For example they are used in MCMC approaches called multi-level algorithms.
The basic idea is to make use of decorrelated and separated domains to improve the statistics of the measurements, e.g.~for two domains the localized parts can be sampled independently which improves the error in the ideal case to $\propto 1/(\sqrt{N}\sqrt{N}) = 1/N$ .

While multi-level or multi-rate sampling are well-known in case of pure gauge simulations, i.e.~they are in general straightfoward to apply
if the action as well as the observable can be decomposed in local terms, it becomes hard if the action as well as the observable is a non-local object. This is in general the case for fermionic contributions depending on the determinant or inverse matrix elements. However, the effective lattice action is local, a necessity for continuum limits, in the sense that separated domain decouples with the distances exponentially proportional to the lowest fermionic mode. This potentially makes multi-level algorithms effective if a suitable decomposition is found.

A successful approach is given by a one-dimensional decomposition in time \cite{Ce:2016idq,Ce:2016ajy}.
Here, the domain is decomposed into two \textit{active} regions, denoted by $\Lambda_0$ and $\Lambda_2$, which are separated by two (one with open boundary conditions) \textit{frozen} domain $\Lambda_1$, see Fig.~\ref{fig:MLdecomposition}.
Now, we can decompose the fermion determinant (here $Q=\gamma_5 D$)
\begin{equation}
 det Q = \frac{ det (1 - \omega )} { det Q_{\Lambda_1} det Q^{-1}_{\Omega_0} det Q^{-1}_{\Omega_2}}  
 \end{equation}
where $Q_{\Lambda_1}$ is defined on the frozen domain $\Lambda_1$,  $Q_{\Omega_0}$ on the domain including $\Lambda_0$ and $\Lambda_1$ and $Q_{\Omega_2}$ on the domain including $\Lambda_2$ and $\Lambda_1$. 
The global term, which includes the contribution between the active domains $\Lambda_0$ and $\Lambda_2$ is given by
\begin{equation}
\omega = P_{\partial \Lambda_0} Q^{-1}_{\Omega_0} Q_{\Lambda_{1,2}} Q^{-1}_{\Omega_2} Q_{\Lambda_{2,0}}
\label{eq:MLgl}
\end{equation}
with the boundary projector $P_{\partial \Lambda_0}$. The global part can be further factorized into a part which can be treated via a multi-boson approach and a global part treated via a reweighting factor, see \cite{Ce:2016ajy}. Note, that fermionic observable can be decomposed in a similar way, see \cite{Ce:2016idq}.

Now, we can define a MCMC method, as follows.
Starting from a thermalized configuration
each active domain is updated via a HMC algorithm localized to $\Omega_0$ and $\Omega_2$ independently from each other $n_1$ times
and on each sample the local part of the observable is measured (note that the global correction enters here as a reweighting factor).
Afterwards by including the global corrections based on eq.~\eqref{eq:MLgl} the global lattice can be updated. At this level also shifts of the lattice are possible.
This is followed by global update steps ideally performed until a new global configuration is obtained. If the procedure is performed $n_0$ times an effective sampling of localized modes with an statistical error $\propto 1/(n_1 \sqrt{n_0})$ is obtained.

\begin{wrapfigure}[17]{R}{0.5\textwidth}
\includegraphics[width=0.5\textwidth]{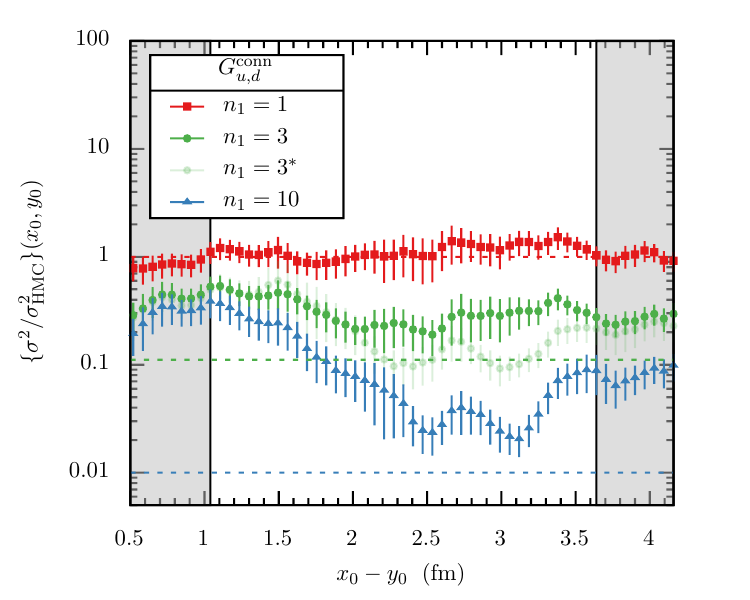}
\caption{\label{fig:MLred}Reduction of the error in $\sigma^2_{G^{conn}_{u,d}}$ using $n_1$ multilevel steps 
with respect the distance $x_0-y_0$ (taken from \cite{DallaBrida:2020cik}).}
\end{wrapfigure}

Alternatively, one can use $n_0$ configurations of an generated ensemble and perform the local updates on each of them. 
This leads to an improved error as demonstrated \cite{DallaBrida:2020cik} in case of the HVP of magnetic moment of the muon on a lattice with $L=48$, $a=0.065 \; \textrm{fm}$ at $300 \; \textrm{MeV}$ pions with domain sizes  $\Lambda_1=8$ and $\Lambda_{0/2}=40$.
In case of the HVP the long distance contribution are notoriously difficult to estimate due to an exponential increase of the noise 
given by
$\frac{\sigma^2_{G^{conn}_{u,d}}(x_0)}{[G^{{conn}_{u,d}}(x_0)]^2} \propto \frac{1}{n_0} e^{2 (M_\rho - M_\pi)|x_0|}$.
By using a statistics of $n_0=25$ configuration separated each by 48 MDUs and $n_1=10$, a precision of the full HVP contribution to $1\%$ could be achieved,
see Fig.~\ref{fig:MLred} for the effective error reductions.

Especially in combination with masterfield simulations
multilevel approaches have the potential to change the way how observables are measured on the lattice.
Despite this, still for many applications the potential is not well understood, e.g.~how effective the method becomes if physical pion masses are approached,
because the required distance between active domain is increased.
Multilevel approaches add an additional level of complexity also on the level of implementations and parallelization.
It is well suited for modular supercomputing and to overcome strong scalability issues, compare \cite{Boyle:2022pai,Boyle:2022ncb},
but an efficient (open source) implementation is so far missing. Python APIs, such as lyncs \cite{Lattice22Bacchio} or GPT \cite{Lattice22Lehner}, are potentially well-suited package,
where the complexity of the multi-level algorithm can be managed on the higher level.

\subsubsection{Gauge updates without topological barriers}
\label{subsec:gupTP}

Let us take a closer look to gauge proposals, which can change topological sectors. As pointed out \cite{Schaefer:2010hu}, topological freezing towards fine
lattices is connected to the gauge group and is also present if fermions are neglected. 
Because of the additional costs, coming with fermions, possible gauge proposals are investigated within pure gauge, before fermions are added.

An idea for a solid gauge proposal is to minimise the number of changed variables or gauge links by allowing possible flips of the topological charge.
A possible transition in two dimensional U(1)-pure gauge theory, if fermions are present also known as Schwinger model, is given by winding the fields 
\cite{Smit:1986fn,Leinweber:2003sj,Durr:2012te,Albandea:2021lvl}
\begin{equation}
U_\mu (x) \rightarrow U^{\Omega}_\mu (x) = \Omega(x) U_\mu(x)  \Omega(x+\hat{\mu})
\label{eq:winds}
\end{equation}
with $\Omega^{\pm}(x_n) = e^{\pm \frac{\pi}{2}(\frac{n}{L_w}+r)}$ on the domain $L_w \times L_w$ with a random shift $r$.
The proposed configuration is then accepted with the probability
$P_{acc}(U\rightarrow U') = \textrm{min}[1 , e^{-S[U']+S[U]}]$.
In combination with a HMC step after each winding step, the algorithm becomes ergodic.

\begin{wrapfigure}[14]{R}{0.4\textwidth}
\includegraphics[width=0.4\textwidth]{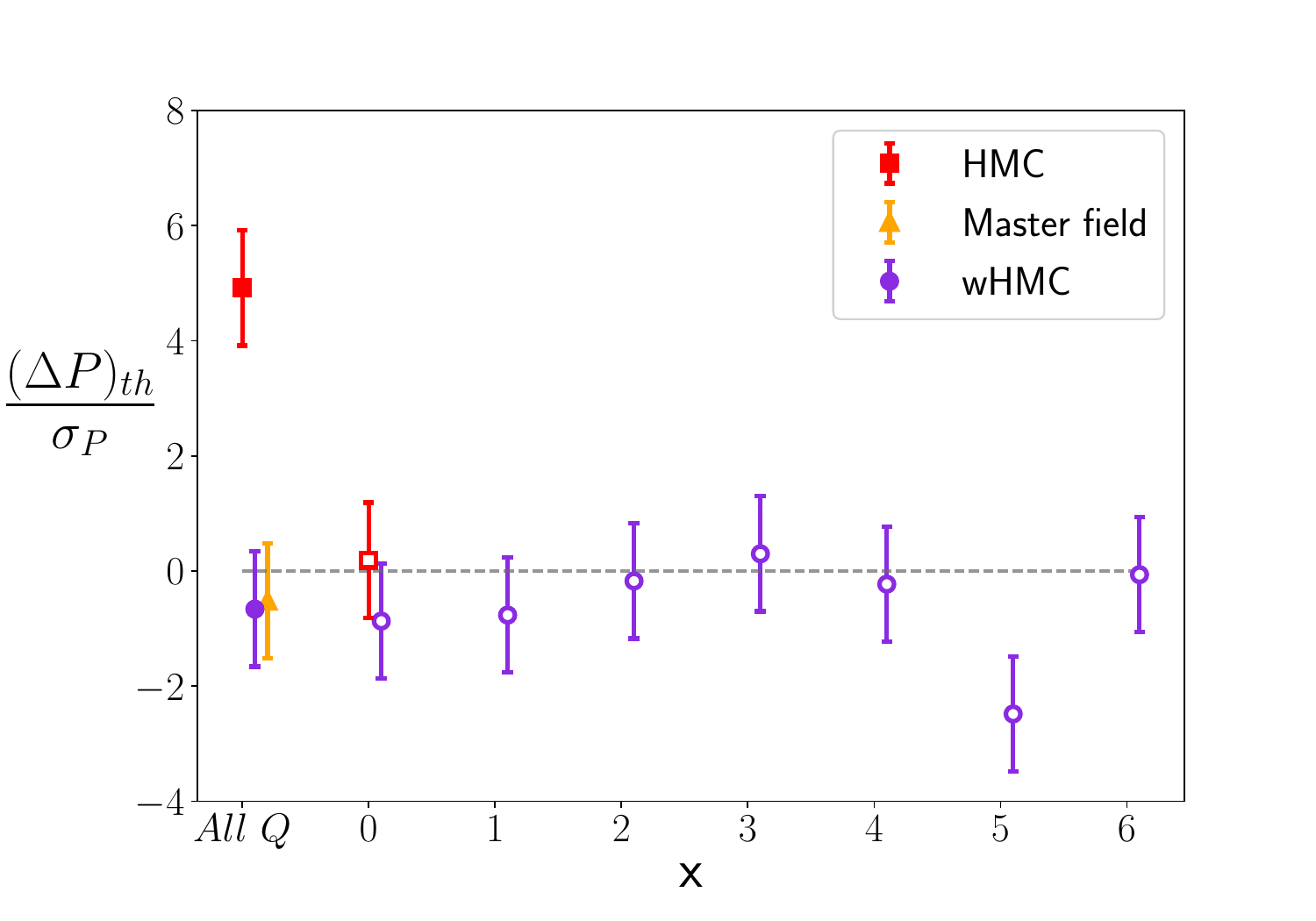}
\caption{\label{fig:quenched_windings}The figure shows the normalized difference between the measured plaquette and the analytical expectation for the different algorithms 
presented in \cite{Albandea:2021lvl}.}
\end{wrapfigure}

This was tested in the 2D Schwinger model at a gauge coupling of $\beta=11.25$
and compared to simulation with a HMC at fix topological charge and a \textit{masterfield} 
simulation using a large volume with lattice extent $L=8192$, see Fig.~\ref{fig:quenched_windings}.

The acceptance rate of the winding proposal might break down towards very fine lattice, i.e.~the phase of the transformation $\Omega^{\pm}(x_n)$ does not depends on the gauge coupling and likely pushes the gaugefield out of the equilibrium for very fine lattice spacings.
Adaptations to implement a similar transformation in SU(3) theories, are so far unsuccessful due to a break down of the acceptance rate, see 
\cite{Eichhorn:2022wxn}.

A more promising approach for updates in SU(N) theories is given by multi-tempering algorithms \cite{Hasenbusch:2017unr,Bonanno:2020hht}, successfully applied in four dimensions.
The idea is to introduce a  defect within the lattice, given by a small cube with links set to zero.
The defect enables smooth transition between topological sector, similar to open boundary.
To recover the target lattice $j=0$, without a defect, a weight parameter $w_j = 1- j/ (N_{lvl}-1)$  is multiplied with the plaquettes located at the defect,
where $N_{lvl}$ is the number of levels.
Now, we can generate a Markov chain by pairing usual update, like a HMC algorithm, with transition accept-reject steps between different level.
If the transition acceptance rate is tuned, which can be done by varying the number of levels, the algorithm can change topological sector smoothly,
as demonstrated in \cite{Bonanno:2020hht,Bonanno:2022yjr}.
Note, that the computational cost are increased with the number of levels, which will increase the cost towards very fine lattices likely significantly.


\subsubsection{Generative models for gauge theories}
\label{subsec:generative}

\begin{figure}
\begin{center}
\includegraphics[width=0.85\textwidth]{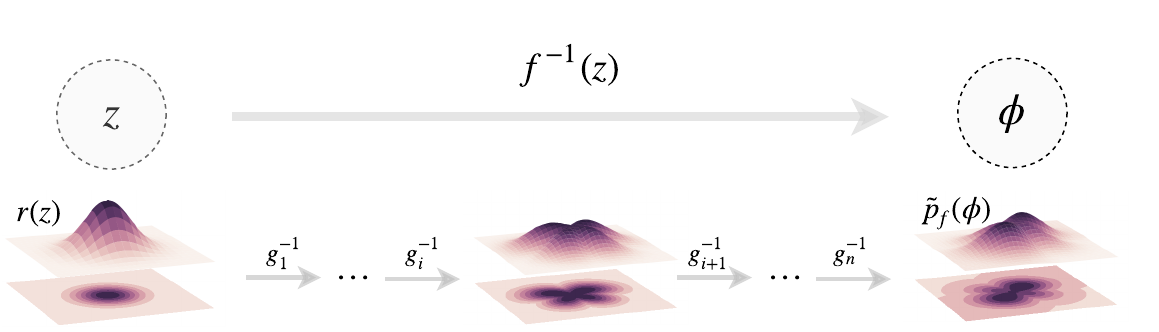}
\caption{\label{fig:flow}The figure shows the construction of the normalized flow using coupling layers $g_i^{-1}$ in case of the $\phi^4$ model 
(taken from \cite{Albergo:2019eim}, note that $\tilde{p}_f(\phi)$ corresponds here to $q(\phi)$). }
\end{center}
\end{figure}

The acceptance rate of the accept-reject step eq.~\eqref{eq:acc} can be also controlled by making use of correlations between $q(U)$ and $p(U)$.
An idea is to use neural networks in combination with machine learning to train the corresponding correlations.
A possible approach is given by generative model, which are applied to pure gauge theories using gauge equivariant flows \cite{Albergo:2019eim,Kanwar:2020xzo,Boyda:2020hsi,Albergo:2021vyo}.

The idea is to use a flow map $f^{-1}(z)$  to propose new configurations with known distribution
\begin{equation}
q(\phi) = r(f(\phi)) \cdot \left| \textrm{det} \frac{\partial f(\phi)}{\partial \phi} \right|~.
\end{equation}
with $r(z)$ a random distribution.
By writing the map $f^{-1}(z)$ as a product over coupling layers 
\begin{equation}
g^{-1}(z) = 
\left\{
\begin{matrix}
 \phi_a = \; z_a \phantom{- t_i(z_a)) \cdot e^{-s_i(z_a)}}\\
 \; \; \phi_b  = (z_b - t_i(z_a)) \cdot e^{-s_i(z_a)}
\end{matrix}
\right.
\end{equation}
which updates a subset of variables $\phi_b$ using only a set of frozen parameters $\phi_a$,
the Jacobian is block-diagonal and is simple to compute. In case of the $\phi^4$ model one can use an even-odd mapping, where all variables at even points belong to one subset and the variables at the odd points to the other subset.
By using gauge invariant objects, e.g.~plaquettes instead of gauge links, as in- and output the coupling layers
becomes gauge equivariant. Links are than updated with the corresponding plaquettes, see \cite{Kanwar:2020xzo}.

The neural networks $s_i$ and $t_i$, e.g.~which consist of convolutional kernels with few hidden layers,
can be trained by minimizing the loss-function 
\begin{equation}
L(q)= D_{KL}(q || p) - \textrm{log}(Z) = \int \prod_j d\phi_j \, q(\phi) (\textrm{log} (q(\phi)) + S(\phi))~.
\end{equation}
This can be done by drawing random samples $z$ and by adjusting the weights in the coupling layers.
One step is done on a set of samples, also called batch, on which the loss-function can be approximated.

After a sufficient number of training steps, e.g.~the effective sample size per configuration $ \textrm{ESS}=1/N (\sum_{i=1}^N p(\phi_i)/q(\phi_i) )^2 /\sum_{i=1}^N (p(\phi_i)/q(\phi_i) )^2  $ reaches a relative high values, the trained gauge equivarient map $f^{-1}(z)$
can be used to draw proposals for gauge configurations, which can be used in MCMC algorithms or in weighted averages.
By combining the proposal with an accept-reject step
$P_{acc} (U\rightarrow U') = \textrm{min} [ 1 \; ,{q(U) p(U')}/{p(U) q(U')}]$
we obtain an exact algorithm, which samples configuration distributed with $p(U)$.

\begin{wrapfigure}[17]{R}{0.5\textwidth}
\vspace{-0.6cm}
\includegraphics[width=0.5\textwidth]{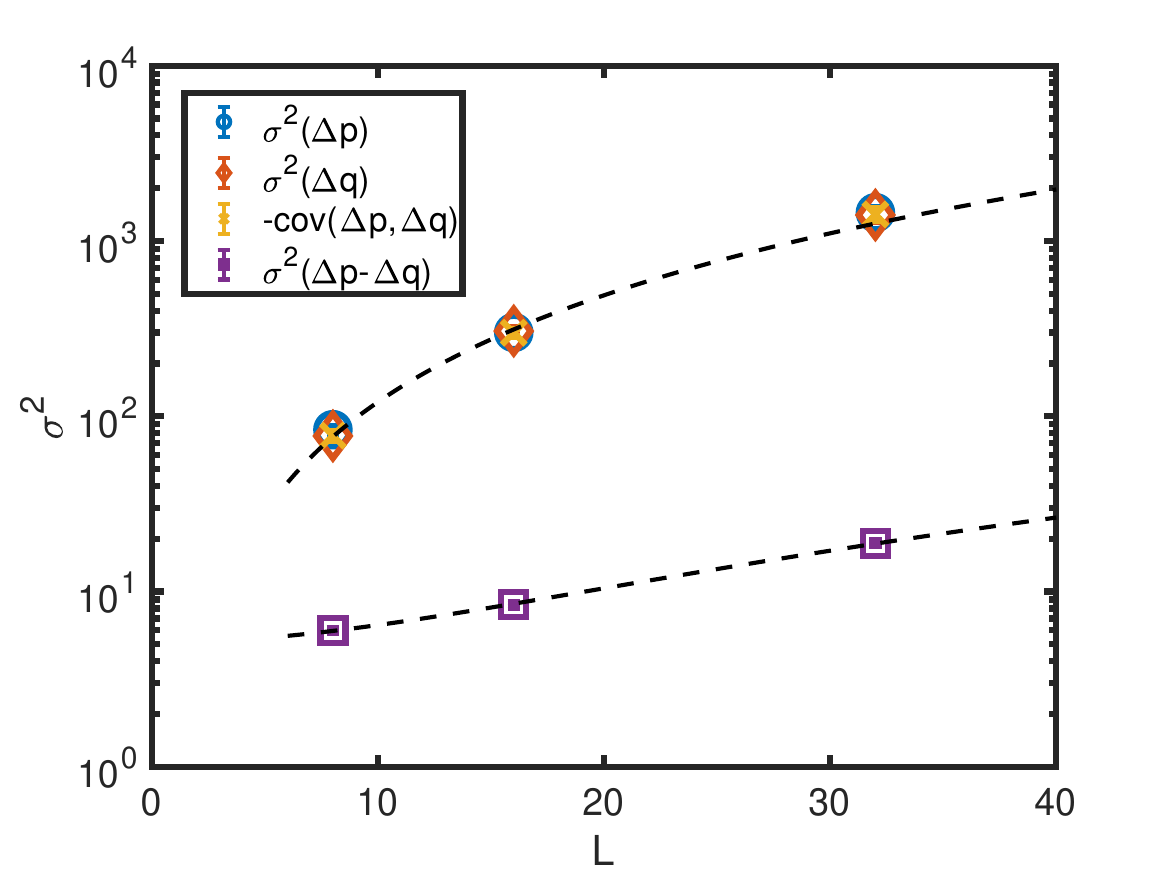}
\caption{\label{fig:finetune}The variances of the log-distribution obtained by training gauge equivariant maps in the 2D U(1) in dependence of the volume is shown.
The number of coupling layers are increased with the lattice extent and the training is stopped after the improvements of the optimizer deterioted.}
\end{wrapfigure}

Generative models are successfully applied to 2D discrete lattice models, such as U(1) theory \cite{Kanwar:2020xzo} or SU(3) \cite{Boyda:2020hsi}. 
Decorrelation between proposed gauges are included in the design, i.e.~they are maximally decorrelated because each initial set is drawn from
a random distribution.
Larger autocorrelation times can only occur due to a small total acceptance rate.



Generative models are a promising novel way to model physics distribution and have the potential to
give new insights into QCD. In contrast to other machine learning applications a
combination with an accept-reject step leads to an exact algorithm.

In case of the 2D-U(1) model, the minimization of the loss function
leads to a minimization of the volume fluctuations, as shown in Fig.~\ref{fig:finetune},
while the overall scaling is still proportional to $\propto V$.
This can be understood as a fine tuning problem.
Let us assume that the distribution are log-normal distributed, than it follows
$\langle P_{acc} \rangle \approx 1 -\frac{\sigma}{\pi}$ for $\sigma \ll 1$
with the variance
$\sigma^2 = \textrm{var}(\Delta p ) + \textrm{var}(\Delta q) + 2 \textrm{cov} (\Delta p, \Delta q)$.
As shown in Fig.~\ref{fig:finetune} we found
that the covariance scales like the variances with $\textrm{var}(\Delta p) + \textrm{var} (\Delta q) \approx -2 \textrm{cov}(\Delta p, \Delta q)$. 
While the major part of the fluctuations  cancels out, the remaining part still increases with the volume.
This illustrate a general problem of the approach, the scalability towards larger volumes,
e.g.~it is not well understood how to reach larger acceptance rates for $L>32$ in case of 2D U(1) model. 
A more complete discussion of the phenomenon can be found in a recent publication \cite{Abbott:2022zsh}.

However this topic, how to achieve scalable generative models, is under active research
and different approaches are considered. Possible research directions are given by investigation of various optimizations
of the maps, e.g.~by modifying neural networks or
using different flows, or by modification of the flow maps, e.g.~using continues flows \cite{Bacchio:2022vje}.

\subsubsection{ Domain decomposed normalizing flows and fermions}

\begin{figure}
\begin{center}
\includegraphics[width=0.95\textwidth]{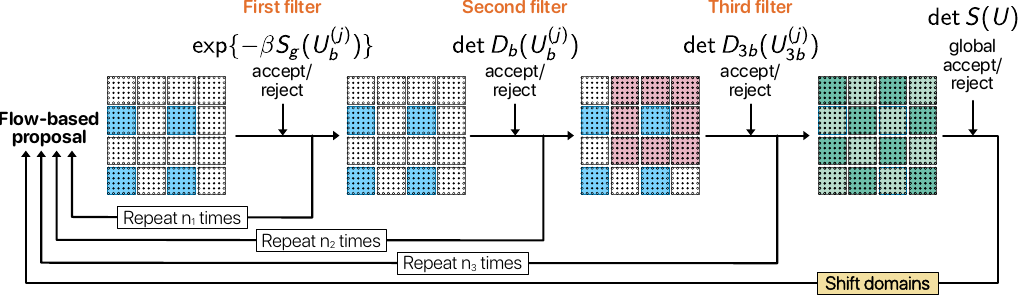}
\caption{\label{fig:flowGC}The steps of the 5 level flowGC algorithm, which iterates starting from the left with a flow-based proposal for the blocks highlighted in blue, followed by nested accept-reject filter steps. In the last step a global correction step is done, followed by a shift of the domains before restarting}
\end{center}
\end{figure}

Another possibility to overcome the volume scaling in generative models is to make use of physical properties of the theory,
in particular by using localization, e.g.~as discussed by factorising the action using domain decomposition.

The idea is to train normalizing flows for local gauge updates within a domain.
This can be done by restricting the flow map and its corresponding coupling layers to variables
within the domain. As shown in the 2D-U(1) model \cite{Finkenrath:2022ogg},
the flow can be trained in a similar fashion to the periodic case \cite{Albergo:2021vyo}.
By freezing boundary terms, updates within a domain are decoupled from the other domains
and each domain can be updated independently, i.e.~updates are independent from the global volume.
Note that to overcome topological freezing the local updates need to enable
topological transition.

To generate configurations with dynamical fermions the fermion determinant has to be included.
This can be done by using accept-reject steps, similar to \cite{Finkenrath:2012az}. 
The general idea is to filter out UV fluctuations step by step using nested hierachical filter steps including correlation between different parts.
Let us write the Boltzmann factor as a product over factorized distributions $p(U) = \prod_j^n P_j(U)$ with the distribution of the $j$th step 
\begin{equation}
 \rho_j(U) = P_0(U,\alpha^{(0)}_i) P_1(U,\alpha^{(1)}_i) \ldots P_j(U,\alpha^{(j)}_i) ~.
\end{equation}
The terms in $p(U)$ can be ordered with respect to the their range, i.e.~$P_0$ contains ultra local actions parts like the pure gauge action, $P_1$ contains local action but short range interactions like block determinants while $P_n$ includes the global Schur complement, which contains long range interactions. 
Now, the accept-reject step of the $j$th-step is given by
 \begin{equation}
P^{j}_{acc} (U\rightarrow U') = \textrm{min} \left[ 1 \;, \frac{\rho_{j-1}(U) \rho_{j}(U')}{\rho_{j-1}(U') \rho_{j}(U)}  \right]~.
\end{equation}
The introduced parameters $\alpha^{(j)}_i$ can be tuned such that the acceptance rate $\langle P^{j}_{acc} \rangle$ starting from $j=n,\ldots,0$ is maximized.

\begin{figure}
\begin{center}
\includegraphics[width=0.475\textwidth]{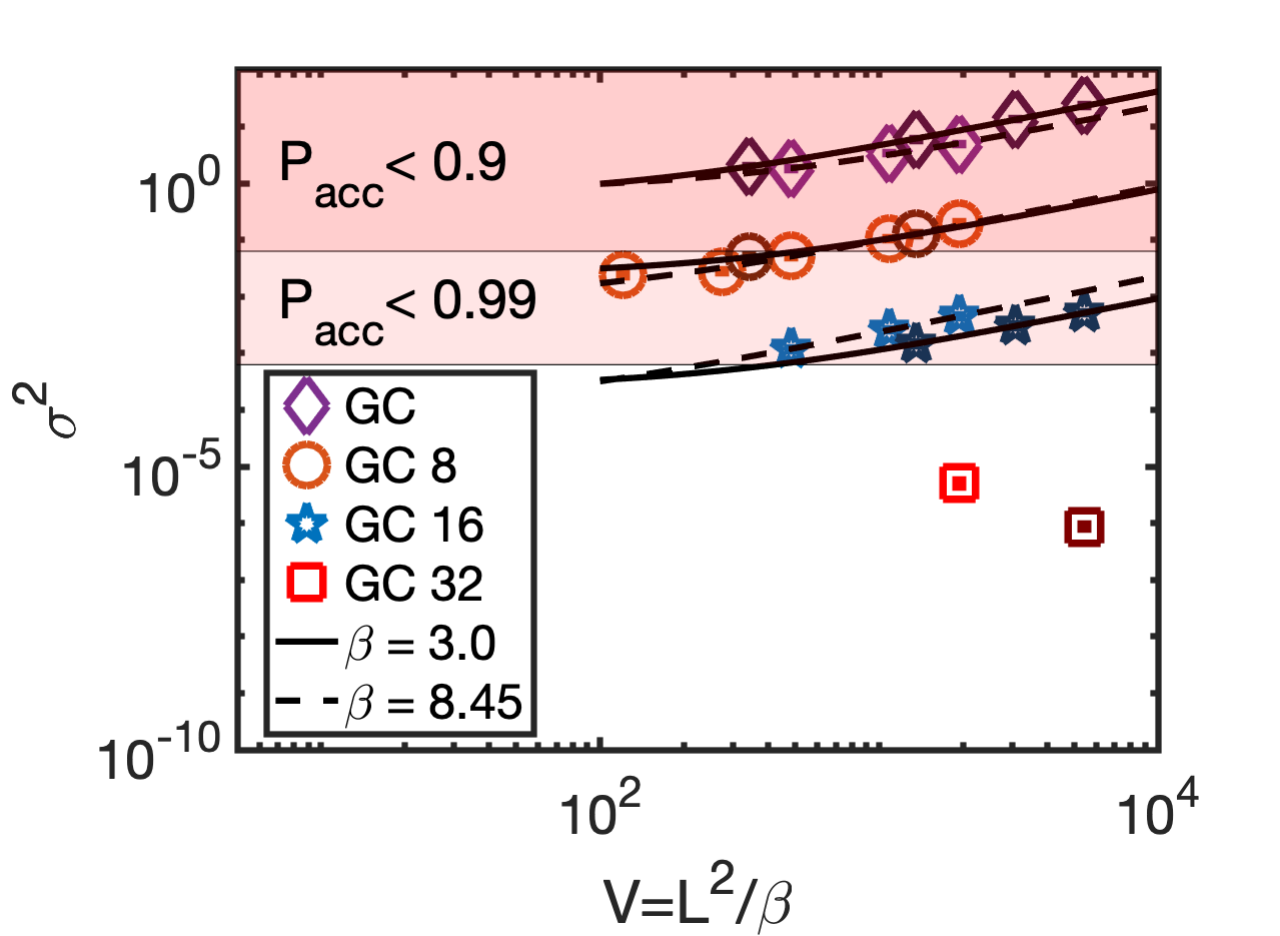}
\includegraphics[width=0.46\textwidth]{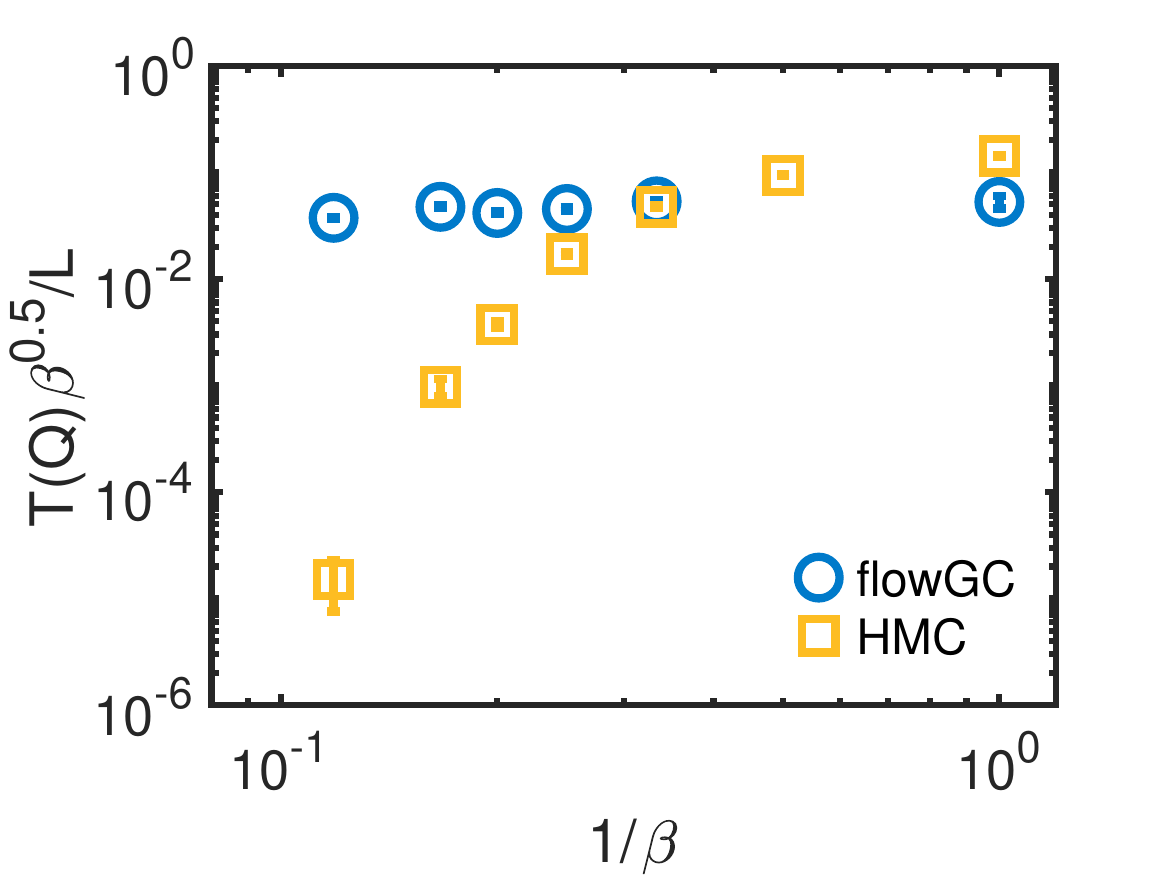}
\caption{\label{fig:SWup}The acceptance rate of the global correction step is shown at constant rate of update links for different distances between active domains is shown in the left figure. On the right side the tunneling rate of the global steps of the flowGC is shown at a constant line of physics with $m_{PS}\sqrt{\beta} = 4$  and $L/\sqrt{\beta} \sim 40$. }
\end{center}
\end{figure}
A MCMC chain is then given by nested hierachical filter steps with gauge updates on domains giving by flow proposals. 
A 4-level approach, denoted as flowGC, is illustrated in Fig.~\ref{fig:flowGC} and is given by 
\begin{itemize}
    \item [0.] Flow proposal to generate $N_0$ samples within each
      active block with lattice extent $l=8$. \vspace{-0.2cm}
    \item [1.] Accept/reject step over the $N_0$ samples using the pure
      gauge action of the active blocks as target probability and
      keeping the final accepted configuration. \vspace{-0.2cm}
    \item [2.] Calculation of the determinant of the block operator
      $D(U_j)$ with $L_b=n\cdot l$ with $n \in 1,2,\ldots$ and accept/reject. Repeat, starting from step 0.,  
      until a sufficient number of links are updated. \vspace{-0.45cm} 
    \item [3.]  \vspace{-0.2cm} Calculation of the determinant of the extended $3L_b \times 3L_b$
      Dirac operator and perform an accept/reject step. Repeat, starting from step 0.,  until a sufficient number of domains are updated. \vspace{-0.2cm}
    \item [4.] Calculation of Schur complement term performing a global
      accept/reject step correcting to the target probability. \vspace{-0.2cm}
\end{itemize}   
This is sufficient to highly suppress fermionic fluctuations, 
if the active domains are separated by a distance of $d=32$ very high acceptance rate were obtained, see left panel of Fig.~\ref{fig:SWup} where the domain extent $l$ is set to the distance of the active domains $d$ to keep the number of active links fixed at different separations $d$ and lattice volumes \cite{Finkenrath:2022ogg}.
By using a domain size of $l=8$ the flowGC can sample sufficiently topological sectors, as depicted on the right panel of Fig.~\ref{fig:SWup},
in case of the 2D Schwinger model at a constant line of physics with $m_{PS}\sqrt{\beta} = 4$  and $L/\sqrt{\beta} \sim 40$
and updates of  $16\%$ of the links.

Towards larger and more complex systems, the second and third filters become potentially a bottleneck by likely developing low acceptance rates.
This effect could be mild down by flow updates which also takes into account factors of the factorised fermion weight.


 
  

\subsubsection{Flows with fermions}
  \begin{wrapfigure}[11]{L}{0.4\textwidth}
\includegraphics[width=0.4\textwidth]{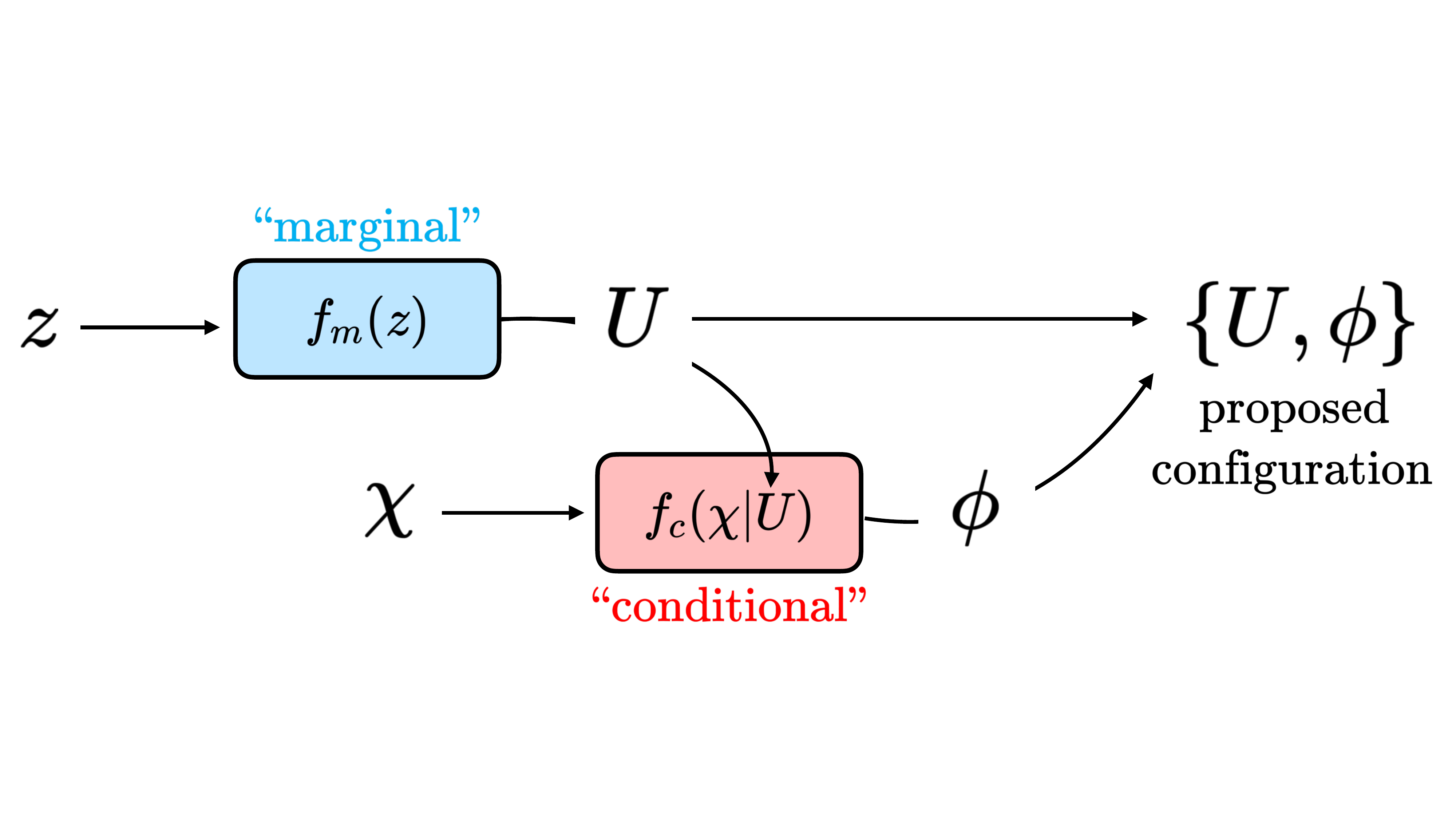}
\caption{\label{fig:diaMLps}The diagram shows the procedure for training flows for pseudofermion and gaugefields (taken from\cite{Abbott:2022zhs}).}
\end{wrapfigure}
The gauge equivariant flow models in subsec.~\ref{subsec:generative} were originally developed for pure gauge systems.
Including the fermion weight increased drastically the computational effort, i.e.~requires frequently computation of the determinant during the training of the weights.
Rewriting the fermion determinant via the pseudofermions integral will decrease computational load to the inverse of the matrix,
however, if treated as a stochastic estimate similar to reweighting \cite{Finkenrath:2013soa}, will add large stochastic fluctuations.
In general this could be tamed by similar methods which increase the accept-reject steps of eq.~\eqref{eq:acc}.

To include fermions within the flows several approaches where investigated, see \cite{Albergo:2021bna}.
A possible way is to enable fermion contributions by sampling pseudofermions \cite{Abbott:2022zhs}.
The general idea is to factorize the distribution into
\begin{equation}
p(U,\phi) = p(U) p(\phi | U) \quad \textrm{with} \quad p(U) \propto \textrm{det} DD^\dagger(U)  e^{-S_g(U)} \quad \textrm{and} \quad p(\phi | U) \propto \frac{e^{-S_{pf}(U,\phi,\phi^\dagger)}}{\textrm{det} DD^\dagger(U)}
\end{equation}
where $p(U)$ is denoted as the marginal distribution and $p(\phi | U)$ the conditional distribution, see Fig.~\ref{fig:diaMLps}.
For the training of the networks the distributions are splited accordingly with $ q(U,\phi)= q(U) q(\phi|U)$.
Now, the idea is to train first the marginal to generate a set of configurations $\{U\}$.
In a second step the pseudofermion map  $f_c(\chi| U)$ is trained on the constant set $\{U\}$.
The flow maps for the pseudofermions require a new design of maps, i.e.~gauge covariance of the networks can be implemented
by using parallel transporter to approximate $D(U)$.

\begin{wrapfigure}[13]{R}{0.4\textwidth}
\includegraphics[width=0.4\textwidth]{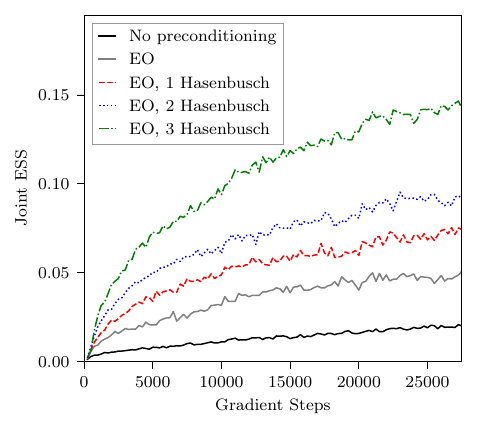}
\caption{\label{fig:ESShasenbusch}The figure shows the effect of using Hasenbusch mass preconditioning in case of generating pseudofermions in case of ESS (taken from \cite{Abbott:2022zhs}).}
\end{wrapfigure}

To further improve the sampling IR/UV-filtering techniques, such as even-odd reductions or Hasenbusch-mass preconditioning, 
can be introduced, i.e~as depicted in Fig.~\ref{fig:ESShasenbusch}.

First results for full QCD in four dimensions with two mass-degenerated fermions 
using the discussed procedures are presented in \cite{Abbott:2022zhs}.
Here, the normalizing flows were used in case of a simulation for a lattice with extent $L=4$
and a mass value of $\kappa=1$  at a coupling constant $\beta=1.0$.

Obviously additional steps are needed in order to apply normalizing flows at the production level in full QCD calculations.
In general this requires a better understanding on how to scale up the systems.
Sampling directly pseudofermion fields can be used to improve other applications, e.g.~in stochastic estimation of the fermions in reweighting~\cite{Finkenrath:2013soa}.

\section{Conclusion}
 
 Advances of algorithms for dynamical fermions in the last decades enable simulation of ensembles at physical pion masses.
To take the next steps towards larger volumes with $L=8 \; \textrm{fm}$, computing time and software package to utilize the pre-exa and exascale machine are
 available although scalability on novel architectures is currently missing.

On the other hand efficient algorithm to unfreeze topology are under development but not available for large lattices, except for using open boundary conditions.
Due to that reaching lattices of size $L = 128$ at fine lattice spacings of  $a = 0.04 \; \textrm{fm}$ need additional algorithmic advances.
Several ideas are investigated where possible solutions could be given by a combination of update steps, which allow transitions between topological sectors,
followed by HMC updates.

To make continuous advances flexible software solutions are required,
in order to develop but also deploy algorithms using efficiently state-of-art HPC systems.
In general high level packages, which can utilize modularity of systems and deal with flexible parallelizations,
can be based on python APIs, such as GPT \cite{Lattice22Lehner} or lyncs \cite{Lattice22Bacchio,Lattice22Yamamoto}, which already enable access to highly optimized lattice kernels
of the packages grid and QUDA, respectively.
This can be very useful for next steps in combining machine learning approaches with lattice QCD algorithms.

To conclude, the community is very active in investigating very diverse approaches and ideas, which will push the frontiers of our understanding of fundamental physics 
in the future.

\textbf{Acknowledgments.}  
The author gratefully thanks the committee of the 39th International Symposium on Lattice Field Theory for the honour 
to give a review on algorithms for dynamical fermions. 
The author thanks for all the effort given to support this work, in particular, special thanks goes to C.~Alexandrou, S.~Bacchio, 
F.~Knechtli and G.~Koutsou.
The author thanks all members of the ETMC for an enjoyable collaboration. 
J.F.~received financial support by the German Research Foundation (DFG) research unit FOR5269 "Future methods for studying confined gluons in QCD", 
by the PRACE Sixth Implementation Phase (PRACE-6IP) program (grant agreement
No.~823767) and by the EuroHPC-JU project EuroCC (grant agreement
No.~951740) of the European Commission. 
Results were obtain on various Supercomputers like Juwels-Booster at JSC, Hawk at HLRS, SuperMUC at LRZ and Frontera at TACC.

\end{document}